\documentclass[a4paper,fleqn,usenatbib,useAMS]{mnras}

\usepackage{graphicx}  
\usepackage{amsmath}  
\usepackage{amssymb}  
\usepackage{multicol}        
\usepackage{bm}    
\usepackage{pdflscape}  
\usepackage{multirow}
\usepackage{xcolor}
\usepackage{hyperref}
\usepackage{booktabs}
\usepackage{subfigure}
\usepackage{xspace}
\usepackage{tikz}
\usepackage[para,online,flushleft]{threeparttable}
\usepackage[colorinlistoftodos,prependcaption,textsize=tiny]{todonotes}
\usepackage{xspace}
\usepackage{orcidlink}
\newcommand{\orcid}[1]{\href{https://orcid.org/#1}{\textcolor[HTML]{A6CE39}{\aiOrcid}}}

\newcommand{\kms}{\xspace\rm\,km\,s^{-1}} 
\newcommand{\hMsun}{\xspace h^{-1}\text{M}_\odot} 
\newcommand{\hkpc}{\xspace h^{-1}\text{kpc}} 
\newcommand{\hmpc}{\xspace h^{-1}\text{Mpc}} 
\newcommand{\rmd}{\mathrm d} 
\newcommand{\ddeg}{^\circ} 
\newcommand{\abacus}{\textsc{ABACUSSUMMIT}\xspace} 

\def\specialname[#1]{\textbf{\textsc{#1}}}

\definecolor{lime}{HTML}{A6CE39}

\usepackage[T1]{fontenc}
\usepackage{ae,aecompl}

\usepackage{newtxtext,newtxmath}

\title[Local Group Analogues in a cosmological context]{
  Local Group analogues in a cosmological context – I. Relating velocity structure to the cosmic web
}
\author[Wang et al.]{
  Kai Wang,$^{1,2}$\thanks{wkcosmology@gmail.com}\orcidlink{0000-0002-3775-0484}
  Peder Norberg,$^{1,2}$\orcidlink{0000-0002-5875-0440}
  Azadeh Fattahi$^{1,3}$\orcidlink{0000-0002-6831-5215}
  and Louis E. Strigari $^{4}$\orcidlink{0000-0001-5672-6079}
  \\
  $^{1}$Institute for Computational Cosmology, Department of Physics, Durham University, South Road, Durham, DH1 3LE, UK\\
  $^{2}$Centre for Extragalactic Astronomy, Department of Physics, Durham University, South Road, Durham DH1 3LE, UK\\
  $^{3}$The Oskar Klein Centre, Department of Physics, Stockholm University, Albanova University Center, 106 91 Stockholm, Sweden\\
  $^{4}$Department of Physics and Astronomy, Mitchell Institute for Fundamental Physics and Astronomy, Texas A\&M University, College Station, Texas, USA
}

\date{Last updated 2025 May 22; in original form 2025 May 5}

\pubyear{2025}

\begin{document}
\label{firstpage}
\pagerange{\pageref{firstpage}--\pageref{lastpage}}
\maketitle


\begin{abstract}
  Our Local Group, dominated in mass by the Milky Way (MW) and M31, provides a unique laboratory for testing $\Lambda$CDM cosmology on small scales owing to its proximity.
  However, its connection to the surrounding large-scale environment, which is essential for interpreting its properties, is inadequately understood.
  In this work, we explore the connection between Local Group analogues (LGAs) and their surrounding large-scale environments using the ABACUSSUMMIT simulation suite, highlighting the key role of the coupling energy of the MW-M31 orbit, $E_{\rm coupling}$.
  We find that LGAs with high $E_{\rm coupling}$ preferentially reside in denser regions, whereas those with low $E_{\rm coupling}$ tend to occupy low-density environments.
  Furthermore, LGAs with low $E_{\rm coupling}$ exhibit strong alignment with cosmic filaments, manifested as a pronounced polar anisotropy in the distribution of tracer haloes.
  By contrast, LGAs with high $E_{\rm coupling}$ show a weaker polar anisotropy but an enhanced azimuthal anisotropy, with large-scale tracer haloes preferentially lying in the plane spanned by the halo pair and the orbital spin vector.
  Within this framework, our Local Group is characterised by typical $E_{\rm coupling}$
  residing in a relatively under-dense environment, yet it remains consistent with the 95\% range of analogue systems identified in the simulation.
\end{abstract}




\section{Introduction}%
\label{sec:introduction}

The Local Group (LG), composed of the Milky Way (MW), Andromeda (M31), and their associated satellite galaxies, is one of the few galaxy systems for which detailed spatial kinematic, and structural properties can be characterised in exceptional details, owing to its proximity \citep[see][for a review]{strigariTimingMassLocal2025}.
The two primary galaxies are each embedded in dark matter haloes with virial masses of $\approx 10^{12}~\rm M_\odot$ \citep[e.g.][]{vandermarelM31TransverseVelocity2008, vandermarelM31VelocityVector2012, vandermarelFirstGaiaDynamics2019, wangMassOurMilky2020}, separated by $\sim 700~\rm kpc$ \citep[e.g.][]{hollandDistanceM31Globular1998, ribasFirstDeterminationDistance2005}, and located in a relatively isolated sector of the cosmic web \citep[e.g.][]{chenELUCIDVICosmic2019, tullyCosmicflows3CosmographyLocal2019}, far from the massive clusters such as Virgo and Fornax.
These well-constrained properties make the LG a uniquely powerful laboratory for testing predictions of the $\Lambda$CDM model on non-linear scales \citep[see][for a review]{bullockSmallScaleChallengesLambda2017}.

The $\Lambda$CDM model has been remarkably successful in reproducing the large-scale distribution of matter, including galaxy clustering statistics \citep[e.g.][]{cole2dFGalaxyRedshift2005, eisensteinDetectionBaryonAcoustic2005}, halo mass function \citep[e.g.][]{allenCosmologicalParametersObservations2011, kravtsovFormationGalaxyClusters2012}, and the morphology of the cosmic web \citep[e.g.][]{tempelDetectingFilamentaryPattern2014}.
On smaller, non-linear scales, however, several properties of the LG have been discussed as potential challenges to the $\Lambda$CDM model.
These include the apparent deficit of luminous satellites compared to the predicted abundance of low-mass subhaloes \citep[the "missing satellite" problem, see][]{klypinWhereAreMissing1999}, which has been replaced in more recent years by the "too-many-satellites" problem with new discoveries of ultra-faint satellites around MW \citep[see][]{grausHowLowDoes2019, santos-santosUnabridgedSatelliteLuminosity2025}; the discrepancy between the inferred halo mass of brightest satellite galaxies and the most massive simulated subhaloes \citep[the "too-big-to-fail" problem, see][]{boylan-kolchinTooBigFail2011}; and the presence of thin and kinematically coherent planes of satellites around both MW and M31 \citep[the
  "satellite-plane" problem, see][]{kroupaGreatDiskMilkyWay2005, ibataVelocityAnticorrelationDiametrically2014, pawlowskiRotationallyStabilizedVPOS2013}.
The "missing satellite" problem is widely considered to result from the baryonic processes that suppress galaxy formation in low-mass haloes due to reionization \citep[e.g.][]{bullockReionizationAbundanceGalactic2000}.
The interpretation of these tensions remains unsettled.
The "too-many-satellite" problem is alleviated by including artificially disrupted subhaloes in simulation due to limited resolution \citep[e.g.][]{santos-santosUnabridgedSatelliteLuminosity2025}.
The "too-big-to-fail" problem could be mitigated by baryonic feedback reducing central density of massive subhaloes, thereby making satellite galaxies more susceptible to tidal stripping and bringing their internal kinematics into better agreement with observations
\citep[e.g.][]{brooksBaryonicSolutionMissing2013, pontzenColdDarkMatter2014, sawalaAPOSTLESimulationsSolutions2016}, or by stochastic variations in satellite accretion histories that leave fewer massive subhaloes in individual systems \citep[e.g.][]{jiangComprehensiveAssessmentToo2015}.
The "satellite-plane" problem has been interpreted as evidence of anisotropic and correlated accretion \citep[e.g.][]{libeskindPlanesSatelliteGalaxies2015}.

Broadly speaking, there are two avenues to reconcile the apparent tensions between LG observations and theoretical predictions: either refining the baryonic processes \citep[e.g.][]{sawalaAPOSTLESimulationsSolutions2016, salesBaryonicSolutionsChallenges2022, santos-santosUnabridgedSatelliteLuminosity2025}, or invoking modifications to the fundamental physics of dark matter nature and cosmology \citep[e.g.][]{vogelsbergerSubhaloesSelfinteractingGalactic2012, anderhaldenHintsNatureDark2013}.
However, certain problems resist both approaches.
The "satellite-plane" problem is a prominent example: no current model modification has succeeded in making such thin and kinematically coherent satellite planes more prevalent in simulations.
Explanations based on analog system variations has been disfavoured at the $\sim 4\sigma$ level \citep[e.g.][]{pawlowskiMilkyWaysDisc2020}.
Here, we put forward an alternative perspective: the large-scale cosmic web environment, a factor that has received inadequate attention in this context, may influence the formation and evolution of LG-like systems and play a significant role in shaping their distinctive properties.

Numerous studies have demonstrated the impact on the galaxy and halo evolution from their surround large-scale environment.
For example, halo spins and shapes exhibit preferential alignment with the surrounding filamentary or sheet-like structures of the cosmic web \citep[e.g.][]{hahnPropertiesDarkMatter2007}, and their satellite subhaloes are also found to be in spatial coherence with the anisotropic environment \citep[e.g.][]{zentnerAnisotropicDistributionGalactic2005, libeskindPlanesSatelliteGalaxies2015}.
Such environmental influences also extend beyond internal halo structure.
\citet{forero-romeroLocalGroupCosmic2015} found that, in paired haloes like our LG, both the orbital angular momentum of the pair and the vector connecting the two haloes are correlated with the surrounding cosmic web anisotropy.
Moreover, constrained-realisation approaches have further highlighted the role of large-scale environment in shaping LG properties \citep[e.g.][]{carlesiConstrainedLocalUniversE2016, sawalaSIBELIUSProjectPluribus2022, mcalpineManticoreLocalClusterCatalogue2025}, reinforcing the view that the LG cannot be interpreted in isolation from its cosmic surroundings, as well as the importance of LCDM initial conditions \citep[e.g.][]{wempeConstrainedCosmologicalSimulations2024a, wempeEffectEnvironmentMass2025}.

Such findings suggest that the large-scale structure can regulate not only the internal properties of haloes, but also the dynamics and spatial configuration of halo pair systems.
In the context of LG analogues, this implies that their orbital configuration and satellite distributions may be partially imprinted by their position and orientation within the cosmic web.
Despite these indications, the relationship between LGA and their large-scale environment has yet to be systematically quantified in a statistically robust manner.
In this work, we address this gap by constructing a large sample of LGA from cosmological N-body simulations, characterizing their environments in terms of both overdensity and anisotropy, and examining correlations with their orbital and kinematic properties.
This approach provides a physically motivated framework for interpreting LG's observed properties in the context of its cosmic web environment.

In this work, we construct a large sample of LGAs in the cosmological N-body simulations of \abacus suite
\citep{garrisonABACUSCosmologicalNbody2021, maksimovaABACUSSUMMITMassiveSet2021} and investigate their connections to the surrounding large-scale structure.
The datasets and the criteria of LGA selection are introduced in \S\,\ref{sec:local_group_and_its_analogues}.
The quantification of the connection between LGAs and large-scale cosmic web is presented in \S\,\ref{sec:relating_local_group_analogs_to_cosmic_web}.
All results are summarized in \S\,\ref{sec:summary}.
Throughout this paper, we assume the cosmological parameters in Planck2018 \citep{planckcollaborationPlanck2018Results2020} following the baseline cosmology in \abacus, where $h = 0.6736$, $\Omega_{\rm m}=0.3137$, and $\Omega_{\Lambda}=0.6862$.

\section{Local Group and its Analogues}%
\label{sec:local_group_and_its_analogues}

\begin{figure*}
  \centering
  \includegraphics[width=1.0\linewidth]{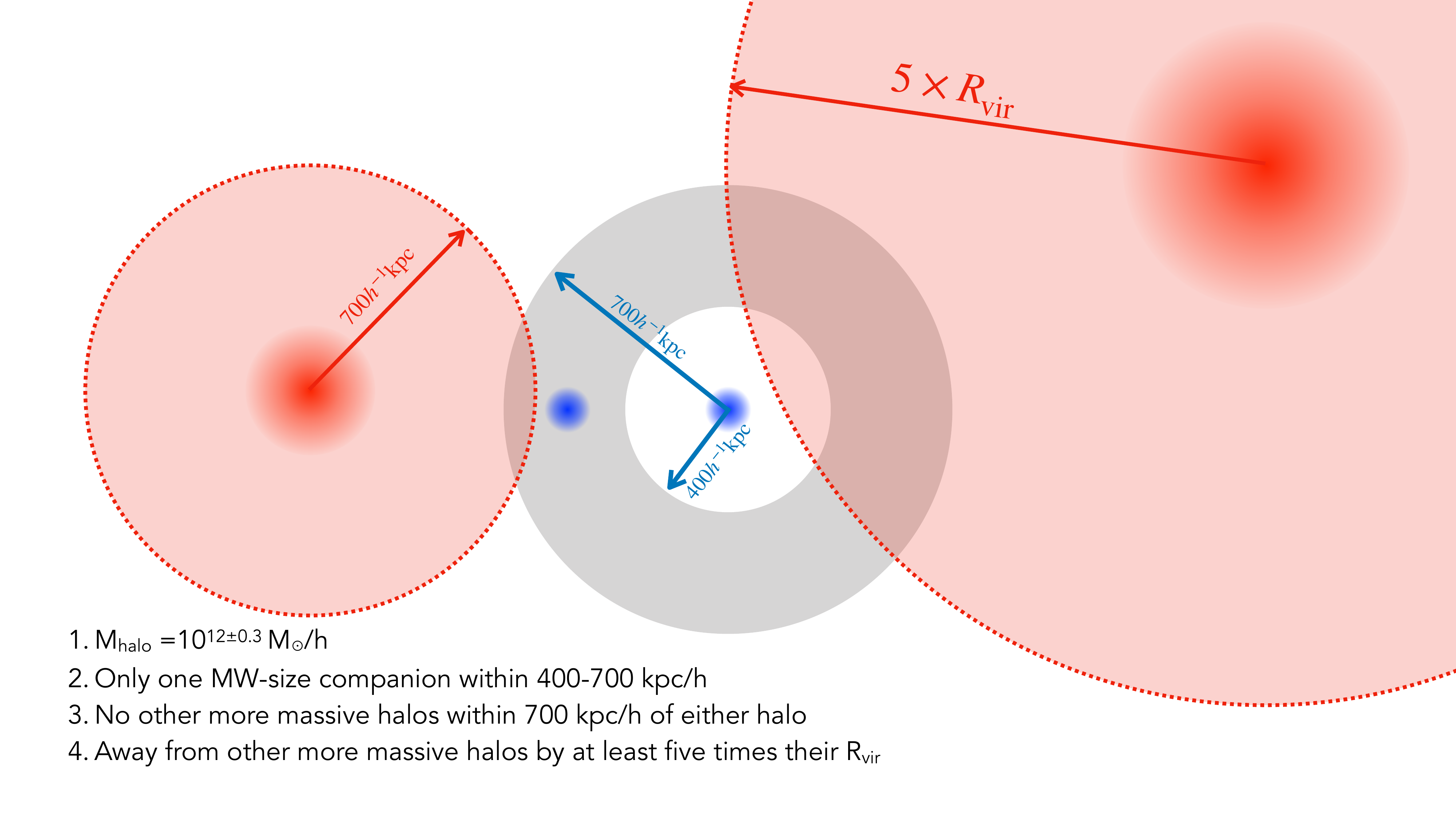}
  \vspace*{-15pt}
  \caption{
    A schematic of the LGA selection criterion.
    A pair of haloes (two blue filled circles) is classified as a LGA if they both satisfy the following four criteria: (1) the virial mass $M_{\rm vir}$ is within $10^{12\pm 0.3}\hMsun$; (2) there is only one halo that satisfy the same mass criterion within $400-700 h^{-1}\rm kpc$ (gray filled band); (3) there are no other haloes above $10^{11.7}h^{-1}\rm M_\odot$ within $700h^{-1}\rm kpc$ of either halo (red filled circle on the left); (4) either halo is away from haloes above $10^{11.7}h^{-1}\rm M_\odot$ (red filled circle ont the right) by at least $5R_{\rm vir}$, where $R_{\rm vir}$ is the virial radius of the neighbor halo.
  }%
  \label{fig:figure/demo}
\end{figure*}

\begin{figure}
  \centering
  \includegraphics[width=1.0\linewidth]{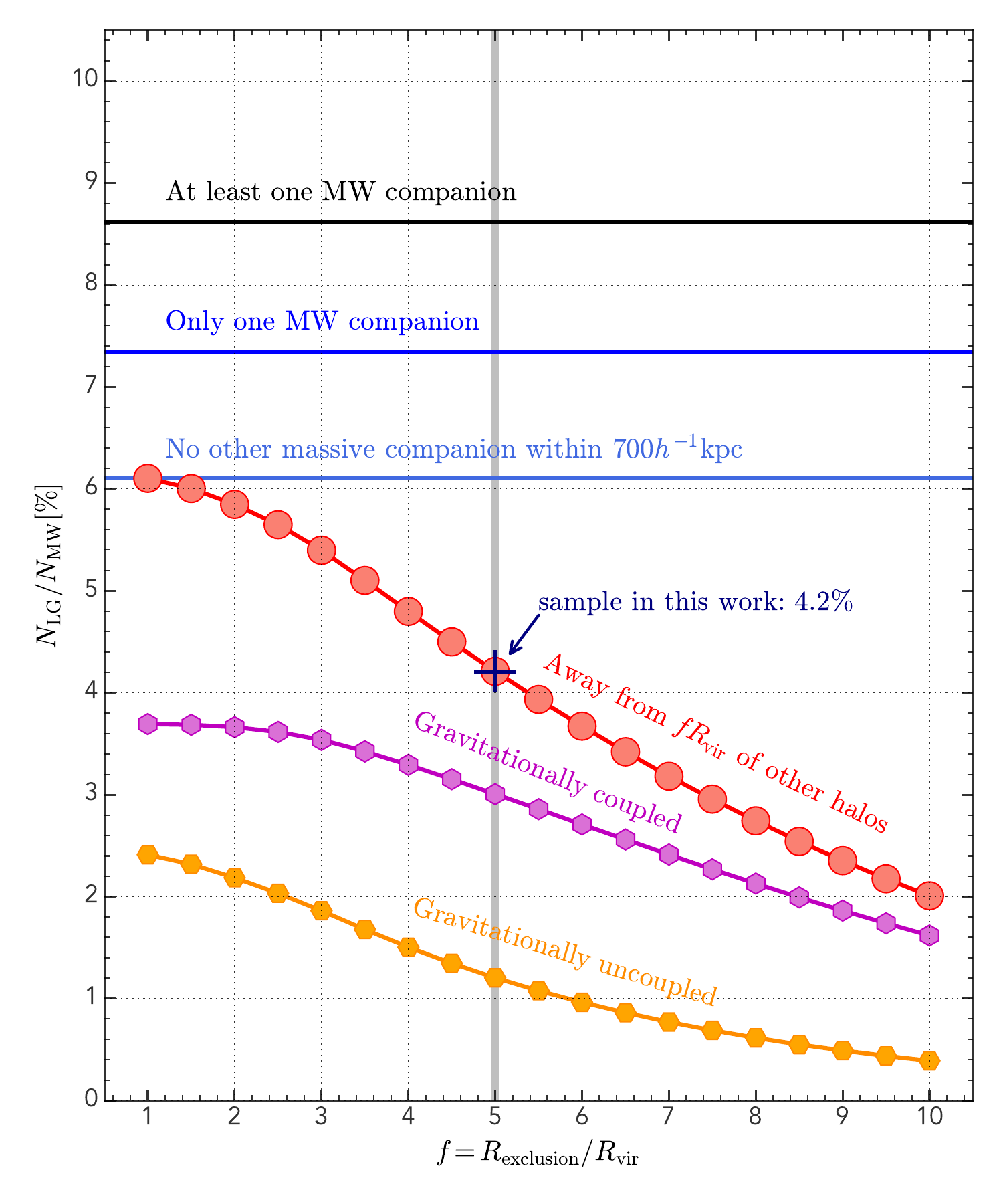}
  \caption{
    The fraction of LGAs among all MW-size haloes ($M_{\rm halo}\in 10^{12\pm 0.3}h^{-1}{\rm M_{\rm \odot}}$) with gradually stricter selection criterion applied.
    The fraction is about 8.6 percent when we require at least one MW-size companion within $400-700h^{-1}{\rm kpc}$ and goes to about 7.4 percent when we require only one MW-size companion.
    The fraction drops to about 6.1 percent when we require no other haloes more massive than
    $10^{12\pm0.3}h^{-1}{\rm M_\odot}$ within $700 h^{-1}{\rm kpc}$ of either halo.
    Finally, we require both haloes to be out of the exclusion region, $R_{\rm exclusion}$, of massive haloes above $10^{12 + 0.3}h^{-1}\rm M_\odot$, where the exclusion region is $fR_{\rm vir}$ for a given dark matter halo, and the fraction of LGAs as a function of $f$ is shown in red circles.
    Here we choose $f=5$ and the eventual fraction is 4.2 percent.
    The magenta and orange pentagons show the fraction of LGAs with $E_{\rm coupling} < 0$ and $E_{\rm coupling} > 0$ (see equation~\eqref{eq:ecoupling}), respectively.
  }%
  \label{fig:figure/lg_fraction_afo_exclusion_radius}
\end{figure}

\begin{figure*}
  \begin{center}
    \includegraphics[width=1.0\linewidth]{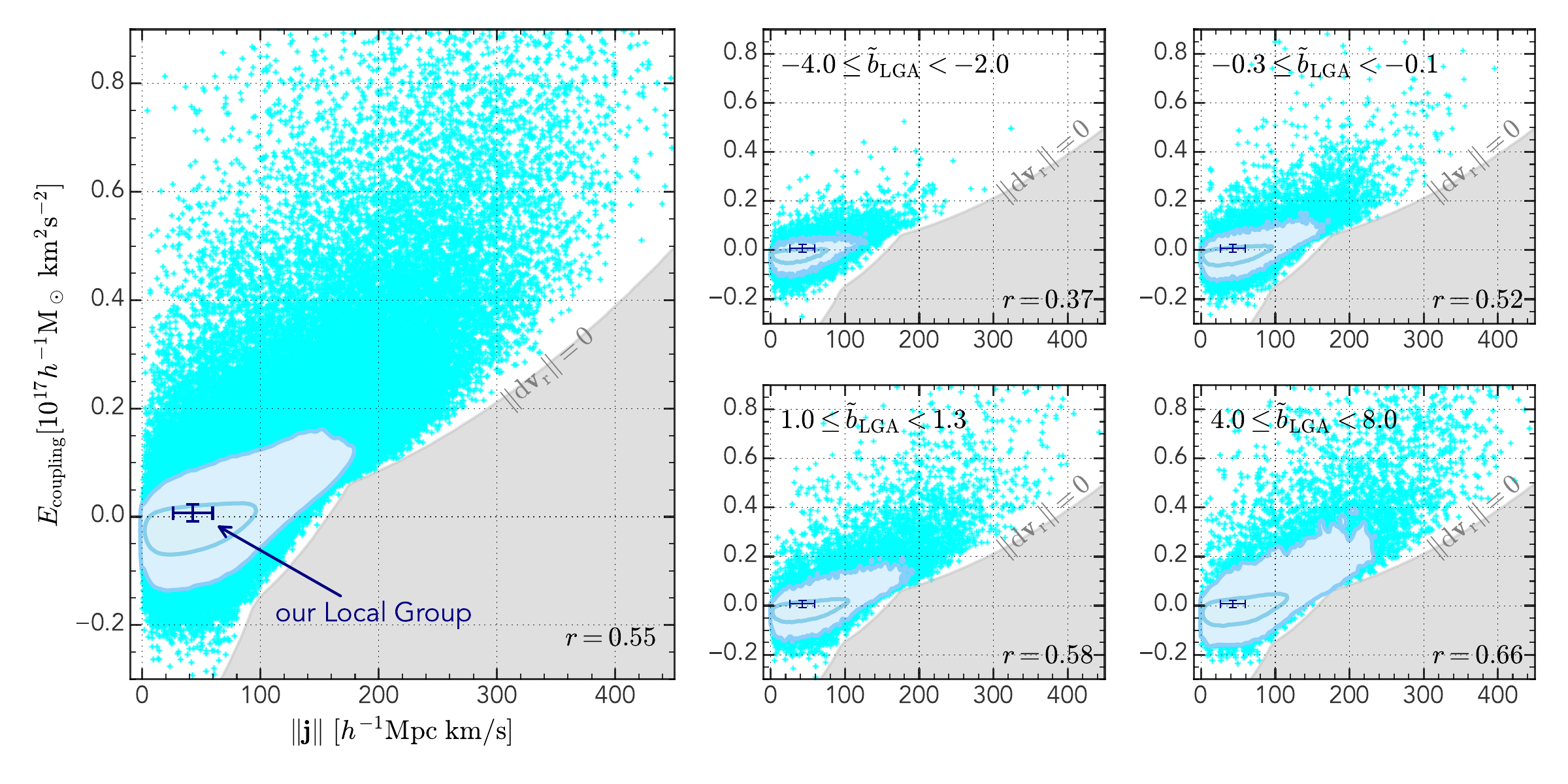}
  \end{center}
  \caption{
  The joint distribution of the coupling energy, $E_{\rm coupling}$, and the specific orbital angular momentum, $\|{\bf j}\|$, for all LGAs (left) and LGAs in different large-scale environmental bins (panels on the right, with the environment range indicated in each panel; see \S\,\ref{sec:large_scale_environment_in_real_space} for the definition of environment).
  In the panels, the contour lines enclose 68 and 95 per cent of the sample and the remaining data points are shown with cyan crosses.
  The bottom right grey region shows the forbidden region separated by $\|\mathrm d\mathbf v_{\rm r}\| = 0$.
  Spearman's rank correlation coefficients between $E_{\rm coupling}$ and $\|\mathbf j\|$ are shown in the lower right corner of each panel for the corresponding sample.
  The result for our LG is presented according to the values in Table~\ref{tab:localgroup}.
  }
  \label{fig:E_J_correlation}
\end{figure*}

\begin{center}
  \begin{table*}
    \caption{Properties of our LG based on measurements in literature
      (see text for the details).}\label{tab:localgroup}
    \begin{threeparttable}[b]
      \renewcommand{\arraystretch}{1.2}
      \begin{tabular}[c]{c|c|c|c|}
        \hline
        Property                                            & Value  & Uncertainty                                & Reference \\
        \hline\hline
        $\log M_{\rm MW}[\hMsun]$                           & 11.96  & 0.1                                        &
        \citet{wangMassOurMilky2020}                                                                                          \\
        \hline
        $\log M_{\rm M31}[\hMsun]$                          & 11.93  & 0.1                                        &
        \citet{vandermarelM31VelocityVector2012}                                                                              \\
        \hline
        $D[\hkpc]$                                          & 519    & 27                                         &
        \citet{vandermarelM31VelocityVector2012}                                                                              \\
        \hline
        $V_{\rm r}[\kms]$                                   & -109.3 & 4.4                                        &
        \citet{vandermarelM31VelocityVector2012}                                                                              \\
        \hline
        $V_{\rm r,pec}[\kms]$                               & -161.2 & 4.7                                        &
        \citet{vandermarelM31VelocityVector2012}                                                                              \\
        \hline
        $V_{\rm t}[\kms]$                                   & 82.4   & 31.2                                       &
        \citet{salomonProperMotionAndromeda2021}                                                                              \\
        \hline
        $E_{\rm coupling}[10^{17}\hMsun {\rm km^2 s^{-2}}]$ & 0.007
                                                            & 0.016  & \S\,\ref{sec:configurations_of_lga_haloes}             \\
        \hline
        $E_{\rm coupling}$ percentile                       & 76.5\%
                                                            & 11.0\% & \S\,\ref{sec:configurations_of_lga_haloes}             \\
        \hline
        $\|{\bf j}\|[\hmpc \kms]$                           & 42     & 16                                         &
        \S\,\ref{sec:configurations_of_lga_haloes}                                                                            \\
        \hline
        $\|{\bf j}\|$ percentile                            & 46.2\% & 19.9\%                                     &
        \S\,\ref{sec:configurations_of_lga_haloes}                                                                            \\
        \hline
        $\Delta_{\rm LG}$                                   & -0.44  & 0.06                                       &
        \S\,\ref{sec:large_scale_environment_in_redshift_space}                                                               \\
        \hline
        $\tilde b_{\rm LG}$                                 & -2.7   & 0.3                                        &
        \S\,\ref{sec:large_scale_environment_in_redshift_space}                                                               \\
        \hline
        $\tilde b_{\rm LG}$ percentile                      & 1.4\%  & 0.3\%                                      &
        \S\,\ref{sec:large_scale_environment_in_redshift_space}                                                               \\
        \hline
      \end{tabular}
      \renewcommand{\arraystretch}{1}
    \end{threeparttable}
  \end{table*}
\end{center}

\subsection{Our Local Group}%
\label{sub:local_group}

Our LG consists of two massive dark matter haloes the Milky Way halo and the M31 halo, separated by $D\approx 770\pm 40\rm kpc$, which is $519\pm 27\hkpc$ \citep[see][]{hollandDistanceM31Globular1998, ribasFirstDeterminationDistance2005, vandermarelM31VelocityVector2012, riessCepheidPeriodLuminosityRelations2012, liSub2DistanceM312021}.
The M31 halo is moving towards the MW halo with radial velocity, $V_{\rm r}\approx -109.3\pm 4.4\,\rm km/s$ \citep[see][]{vandermarelM31VelocityVector2012}, so the radial peculiar velocity is $V_{\rm r, pec}\approx V_{\rm r} - HD = -161.2 \pm 4.7\, \rm km/s$, where $H=h\times 100\,\rm km/s/Mpc$ is the Hubble constant.
The tangential velocity is more uncertain because of the challenge of measuring the proper motion and selecting the tracers.
\citet{vandermarelM31TransverseVelocity2008} obtained the tangential velocity of $42\pm 14\kms$ using M31's satellite galaxies as tracers: \citet{vandermarelM31VelocityVector2012} reported a new tangential velocity measurement of $17.0\pm 17.3\kms$ using the proper motion measurement of M31 stars with \textit{Hubble Space Telescope} after correcting for internal motions.
Then, this value is updated to $57^{+35}_{-31}\kms$ in \citet{vandermarelFirstGaiaDynamics2019} using the \textit{Gaia} Data Release 2 (DR2) and to $82.4\pm 31.2\kms$ in \citet{salomonProperMotionAndromeda2021} based on the Early Third Data Release of the \textit{Gaia} mission (EDR3).
Although the results estimated from \textit{Gaia} DR2 and EDR3 differ by about $40\%$, they are still consistent with each other within the uncertainties quoted.
In this work, we adopt the value from the latest \textit{Gaia} EDR3 measurement in \citet{salomonProperMotionAndromeda2021}.

The halo masses of the Milky Way and M31 systems are constrained by a variety of methods \citep[see][for a review]{wangMassOurMilky2020}.
The inferred values depend not only on the tracer and the method adopted but also on the assumed halo profile and mass definition.
\citet{wangMassOurMilky2020} compiled 11 recent Milky Way halo mass measurements based on the \textit{Gaia} DR2 data and found that the median value of these measurements is about $10^{11.9+/- 0.3}\hMsun$ when the halo mass is defined so that the mean density within a radius just exceeds 200 times the critical density.
This corresponds to $10^{12.0+/- 0.3}\hMsun$ using the halo mass definition in ABACUSSUMMIT (see~\S\,\ref{sub:abacussummit}), assuming a Navarro-Frenk-White mass profile with a concentration of 10 at $z=0$ \citep{navarroUniversalDensityProfile1997, klypinLCDMbasedModelsMilky2002}.
The mass of the M31 halo is estimated to be $10^{11.9\pm 0.4}\hMsun$ based on the timing argument \citep[see][]{vandermarelM31VelocityVector2012}.

\subsection{ABACUSSUMMIT}
\label{sub:abacussummit} 

The ABACUSSUMMIT \citep{garrisonABACUSCosmologicalNbody2021, maksimovaABACUSSUMMITMassiveSet2021} is a suite of cosmological simulations for a broad range of scientific applications.
ABACUSSUMMIT begins from a large volume ($400^3h^{-3}\text{Gpc}^3$) of high accuracy simulation assuming Planck 2018 cosmological parameters, separated into 25 base simulations with each of the box side length $2h^{-1}\text{Gpc}$.
Each base simulation contains $6912^3$
particles with a particle mass of $2\times 10^9h^{-1}\text{M}_\odot$ and force softening of $7.2h^{-1}$ proper kpc.
We use the baseline simulation suite, which considers the Planck2018 parameters, where $\Omega_{\rm m, 0}=0.3137$ and
$h=0.6736$.

Dark matter haloes are identified using the COMPASO algorithm \citep[see][for details]{hadzhiyskaCOMPASONewHalo2022}.
This is a modified Friends-of-Friends algorithm, in which the linking-length is 0.25 times the mean interparticle separation, but only for particles with a kernel density estimate greater than $60$.
The kernel is chosen to be $1 - r^2/b_{\rm kernel}^2$, where $b_{\rm kernel}$ is 0.4 times the mean interparticle separation.
The particle groups from this modified FoF algorithm are referred to as the L0 haloes.
Within each L0 halo, L1 haloes, corresponding to conventional dark matter haloes, are identified using a competitive spherical overdensity algorithm that grows spherical regions around local density maxima.
Finally, within each L1 halo, a secondary spherical search identifies L2 haloes, which trace dense halo cores or prominent substructures.
In this study we use only L1 haloes.
Halo masses are defined as the total mass enclosed within the radius at which the mean density just exceeds $200\rho_{\rm crit}\times [(18\pi^2 + 82x - 39x^2) / (18\pi^2)]$, where $x\equiv \Omega_{\rm m}(z) - 1$ \citep{bryanStatisticalPropertiesXRay1998}.
With this definition, the mean density of L1 haloes is approximately 116, 126, and 135 times the critical density at redshifts of 0, 0.1, and 0.2, respectively, where $\Omega_{\rm m}$ is 0.3137, 0.3783, and 0.4413.
We denote the halo mass defined in this way as $M_{\rm vir}$, and we have $M_{\rm 200c}/M_{\rm vir}=$0.878, 0.896, and 0.912 at the redshift of 0, 0.1, and 0.2 assuming the halo follows the NFW profile with a concentration of 10 \citep{klypinLCDMbasedModelsMilky2002}, where $M_{\rm 200c}$ is the halo mass defined such that the mean density within the virial radius is 200 times the critical density.
The position and velocity of each L1 halo are defined as the centre-of-mass position and velocity of the most massive L2 halo contained within it.

The original COMPASO algorithm finds an excess of low-mass dark matter haloes floating around massive haloes, which results in excessive clustering signal at $1-3h^{-1}\text{Mpc}$.
This artifact has been corrected by applying the cleaning process described in \citet{boseConstructingHighfidelityHalo2022}.
We check the halo pair numbers within the mass range of $10^{12\pm 0.3}\hMsun$ as a function of separation in Appendix\,\ref{sec:halo_pair_number_distribution}: our minimal selection separation ($400h^{-1}\text{kpc}$, see~\S\,\ref{ssub:lga_selection}) is greater than the threshold where the halo pair counts become incomplete.

It is noteworthy that the ABACUSSUMMIT simulation only evolves to $z=0.1$.
Hence we estimate the evolution of the halo mass between $z=0$ and $z=0.1$ using the halo mass accretion rate in \citet{mcbrideMassAccretionRates2009} and the resultant difference is $\lesssim 0.05$ dex for $10^{12}\hMsun$ in our cosmology, which is neglected here.

\subsection{Local Group Analogues}%
\label{sub:local_group_analogues_lga_}

\subsubsection{LGA selection}%
\label{ssub:lga_selection}

We extracted three main features for our LG: both MW and M31 haloes have a mass of about $10^{12}~\hMsun$ \citep{vandermarelM31VelocityVector2012, vandermarelFirstGaiaDynamics2019, wangMassOurMilky2020}; they are separated by about $500~\hkpc$ \citep{mcconnachieDeterminingLocationTip2004}; and they live in a rather isolated environment\footnote{The Virgo cluster is about $11\hmpc$ away and the Fornax cluster is about $13\hmpc$ away.} \citep{chenELUCIDVICosmic2019, tullyCosmicflows3CosmographyLocal2019}.
In order to mimic these three features, we designed four criteria (illustrated in Fig.~\ref{fig:figure/demo}) to select LGAs in the ABACUSSUMMIT simulations:
\begin{enumerate}

  \item
        The virial mass $M_{\rm vir}$ of each halo in the halo pair is within $10^{12\pm 0.3}~\hMsun$. We refer to dark matter haloes within this mass range as MW-mass haloes.

  \item
        The distance between the two haloes in the halo pair is between $400~\hkpc$ and $700~\hkpc$.

  \item
        There are no other haloes more massive than $10^{11.7}~\hMsun$ within $700~\hkpc$ of either halo.

  \item
        Both haloes are at least $R_{\rm exclusion}=5R_{\rm vir}$ away from other haloes with mass above $10^{12.3}~\hMsun$, where $R_{\rm vir}$ is the virial radius of the other halo\footnote{This environmental criterion is imposed since previous studies found that massive haloes can impact galaxy properties out to about $5R_{\rm vir}$ \citep{wangDissectTwohaloGalactic2023}.}.

\end{enumerate}

The resultant fraction of LGAs among all MW-mass haloes under progressively stricter selection criteria is shown in Fig.~\ref{fig:figure/lg_fraction_afo_exclusion_radius}.
Approximately 8.6\% of MW-mass haloes have at least one MW-mass companion. Among these, $\approx 7.4\%$ has exactly one companion, implying that the remaining $\approx 1.2\%$ are systems with two or more companions, corresponding to triplets or more complex configurations.
The second isolation criterion further decreases the fraction to $\approx 4.2\%$, which constitutes the LGA sample used for this study.

Each LGA system contains two haloes that are treated symmetrically in our analysis. To facilitate systematic studies, we artificially designate one halo as the primary halo (MW analogue) and the remaining one as the secondary halo (M31 analogue), then reverse these roles.
This bidirectional treatment means that each physical halo pair generates two distinct entries in our analysis catalog.

The selection criteria of LGAs varies significantly across studies, leading to considerable differences in sample sizes \citep[e.g.][]{liMassesLocalGroup2008, fattahiAPOSTLEProjectLocal2016, hartlLocalGroupTiming2022, sawalaLocalGroupsMass2023}.
For instance, the number of LGAs selected in the Millennium simulation with galaxies populated with the semi-analytical model varies from $\sim 100$ to $\sim 10,000$, depending on mass thresholds and morphological cuts \citep[e.g.][]{liMassesLocalGroup2008}.
Most selection criteria focus on three key system properties, which are mass, separation, and isolation, as adopted in our work.
Some studies impose additional relative velocity constraints: \citet{hartlLocalGroupTiming2022} requires negative radial relative velocity to ensure the two haloes are approaching, \citet{sawalaLocalGroupsMass2023} imposes a velocity upper limit to both the radial and tangential components, and \citet{fattahiAPOSTLEProjectLocal2016} imposed additional requirements on the recession velocities of outer LG members.
Nevertheless, only $\approx 7\%$ LGA systems in our selection have positive radial velocity (see \S\,\ref{sec:relative_velocity_and_large_scale_environment} for details), thus these velocity constraints only affect a very small fraction of potential LGAs, which justifies our decision to omit them.

\subsubsection{Configurations of LGA haloes}
\label{sec:configurations_of_lga_haloes} 

For each LGA system, we can define the displacement vector, $\mathrm d\mathbf r$, starting from the primary halo to the secondary halo, and the relative velocity vector, $\mathrm d\mathbf v\equiv \mathbf v_{\rm s} - \mathbf v_{\rm p}$, as the velocity difference between the secondary ($\mathbf v_{\rm s}$) and primary ($\mathbf v_{\rm p}$) haloes.
Then, we can decompose the relative velocity into a radial component\footnote{Here we use the relative peculiar velocity without taking into account the Hubble-flow contribution to the pairwise motion.} and a tangential component, as
\begin{equation}
  \mathrm d\mathbf v_{\rm r} \equiv \frac{(\mathrm d\mathbf v\cdot\mathrm d\mathbf r)\mathrm d\mathbf r}{\|\mathrm d\mathbf r\|^2} \label{eq:vr}
\end{equation}
and
\begin{equation}
  \mathrm d\mathbf v_{\rm t} \equiv \mathrm d\mathbf v - \mathrm d\mathbf v_{\rm r}
  \label{eq:vt}
\end{equation}
The specific orbital angular momentum is defined as
\begin{equation}
  \mathbf j \equiv \mathrm d\mathbf r\times \mathrm d\mathbf v
  \label{eq:ang_momentum}
\end{equation}
which amplitude is $\|\mathbf j\|= \|\mathrm d\mathbf r\| \|\mathrm d\mathbf v_{\rm t}\|$.
Finally, we define the coupling energy as
\begin{equation}
  E_{\rm coupling} = \frac12 \frac{M_{\rm p}M_{\rm s}}{M_{\rm p} + M_{\rm s}}\|\mathrm d\mathbf v\|^2 - \frac{GM_{\rm p}M_{\rm s}}{\|\mathrm d\mathbf r\|}
  \label{eq:ecoupling}
\end{equation}
where $M$ denotes the halo mass, and the suffixes `p' and `s' are for `primary' and `secondary', respectively.
Here the coupling energy is defined as the total energy of the LGA system, neglecting the spatial extension of each halo.
The sign of $E_{\rm coupling}$ roughly indicates whether the two haloes are coupled ($E_{\rm coupling} < 0$) and will likely remain associated, or uncoupled ($E_{\rm coupling} > 0$) and expected to separate during subsequent evolution.
As shown in Fig.~\ref{fig:figure/lg_fraction_afo_exclusion_radius}, the fraction of both coupled and uncoupled LGAs decreases with increasing exclusion radius, while coupled LGAs become increasingly dominant within the LGA population at larger exclusion radii. For our fiducial choice $(R_{\rm exclusion}/R_{\rm vir}=5)$, coupled and uncoupled LGAs account for about 3.0\% and 1.2\% of all MW haloes, respectively.
Together, these systems define the LGA sample in this study, comprising about $4.2\%$ of the full MW-mass halo population.

Fig.~\ref{fig:E_J_correlation} shows the joint distribution of the coupling energy, $E_{\rm coupling}$, and the amplitude of the specific orbital angular momentum, $\|\mathbf j\|$, for all LGAs in the larger panel and in different large-scale environment bins in smaller panels on the side.
These two quantities are strongly correlated to each other, where LGAs with high specific orbital angular momentum also have higher coupling energy.
The strength of this correlation is quantified by Spearman's correlation coefficient, $r$, as indicated in the lower right corner of each panel.
Moreover, the correlation is strengthened as the large-scale overdensity (see \S\,\ref{sec:large_scale_environment_in_real_space} for the definition of environment) increases, where Spearman's correlation coefficient increases from $\approx 0.37$ in the lowest-density bin to $\approx 0.66$ in the highest-density bin.
Finally, the gray shaded region shows the forbidden region separated by the $\|\mathrm d\mathbf v_{\rm r}\| = 0$ line which defines the minimal coupling energy for a given specific orbital angular momentum amplitude.

Based on the observed proper motion measurement of M31 in
\citet{salomonProperMotionAndromeda2021}, we estimate the amplitude of the specific angular momentum of the LG to be
\begin{equation}
  \|\mathbf j\|_{\rm LG} = (42\pm 16) \hmpc \kms
\end{equation}
and the coupling energy\footnote{We use the peculiar relative velocity to compute the coupling energy in equation~\eqref{eq:ecoupling}. If instead one uses the physical relative velocity between the MW and M31, i.e. including the Hubble-flow contribution, the inferred coupling energy becomes negative.} to be
\begin{equation}
  E_{\rm coupling, LG} = (0.007\pm 0.016) \times 10^{17}\hMsun {\rm km^2s^{-2}}.
\end{equation}
The uncertainty is estimated as the standard deviation of 10,000 realizations with all of the measured quantities sampled from their own distributions, assumed to be independent Gaussian distributions.
The location of our LG on the $E_{\rm coupling}-\|\mathbf j\|$ plane is shown in Fig.~\ref{fig:E_J_correlation}: our Local Group lies close to the centre of the joint distribution, indicating means that the LGAs identified in the simulation naturally encompass the observed values of $E_{\rm coupling}$ and $\|\mathbf j\|$ of our Local Group, even though no velocity-based selection was imposed.

\subsubsection{Large-scale environment in real space}
\label{sec:large_scale_environment_in_real_space} 

\begin{figure}
  \centering
  \includegraphics[width=0.9\linewidth]{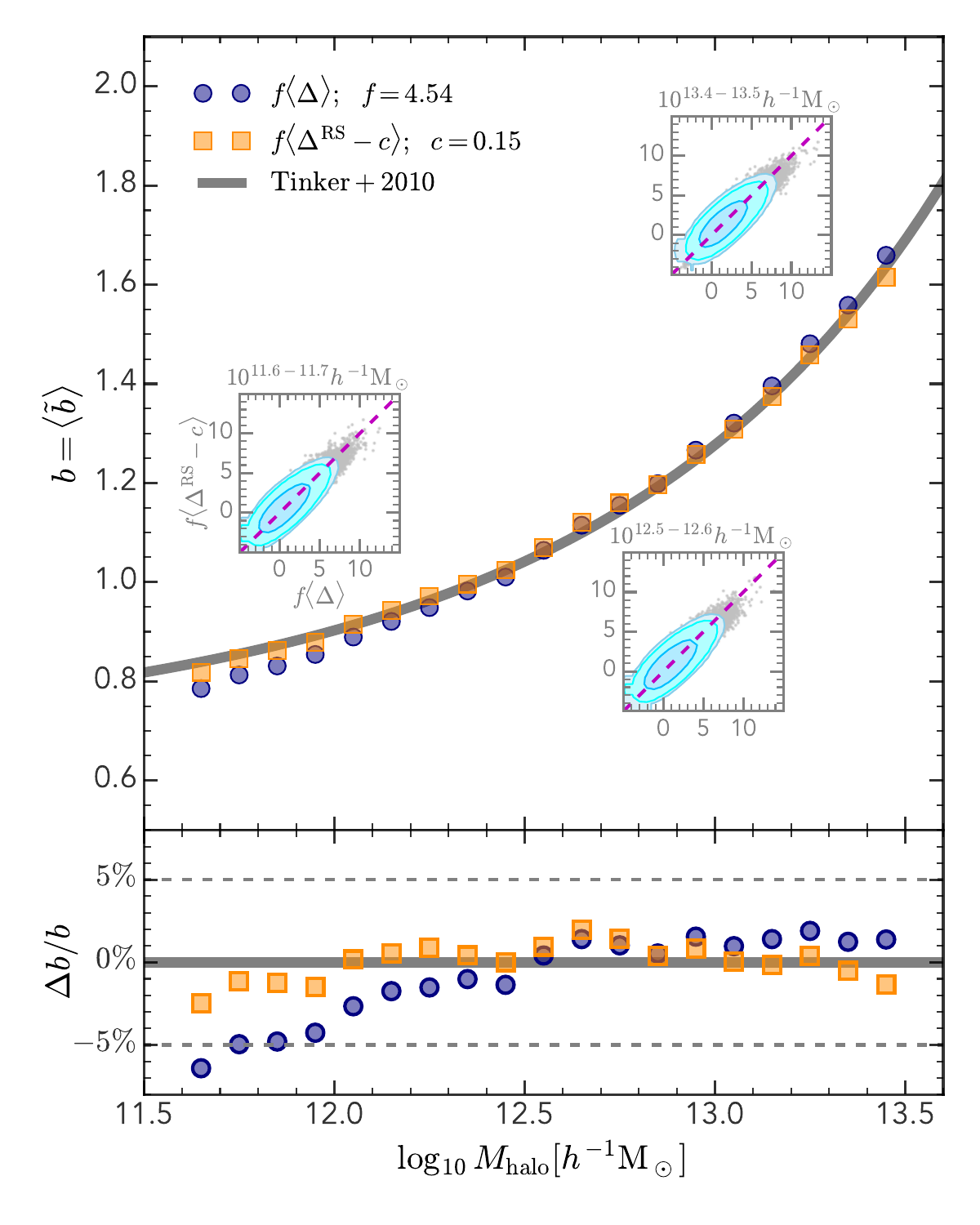}
  \caption{
    \textbf{Top panel:} The blue circles and orange squares show $f\langle
      \Delta\rangle$ and $f\langle \Delta^{\rm RS} - c\rangle$ as a function of halo mass, where $\Delta$ and $\Delta^{\rm RS}$ are the number overdensity of tracer haloes within $5-15\hmpc$ (see equations~\eqref{eq:delta} and \ref{eq:delta_rs}), and $\langle\cdot\rangle$ means the mean value over all haloes in the corresponding halo mass bin.
    The halo bias relation in \citet{tinkerLargescaleBiasDark2010} is shown in gray solid curve.
    The small embedded panels show the comparison between $f\langle \Delta\rangle$ and $f\langle \Delta^{\rm RS} - c\rangle$ in selected halo mass bins.
    \textbf{Bottom panel:} The fractional residual of $f\langle \Delta \rangle$ and $f\langle \Delta^{\rm RS} - c\rangle$ with respect to the halo bias relation in \citet{tinkerLargescaleBiasDark2010} as a function of halo mass.
    Halo bias scales linearly with the mean overdensity of tracer haloes at $5-15\hmpc$, while an additional constant shift is required for the mean overdensity measured in the redshift space.
  }%
  \label{fig:figure/halo_bias}
\end{figure}

In this work, the large-scale environment is quantified by the number of dark matter haloes with mass greater than $10^{11.6}\hMsun$ and within a spherical shell of inner radius $5~\hmpc$ and outer radius $15~\hmpc$, centred on the primary halo.
This mass and distance range is chosen to match the accessible halo population in the Local Volume in the 2MASS Redshift Survey \citep[2MRS;][]{lavaux2MGalaxyRedshift2011, limGalaxyGroupsLowredshift2017}, while excluding regions within $5\hmpc$ to minimize the impact of the isolation criterion used in the LGA selection.

In real space, the number overdensity of tracer haloes is defined as
\begin{equation}
  \Delta \equiv \frac{n - \bar n}{\bar n}
  \label{eq:delta}
\end{equation}
where $n$ is the number density of tracer haloes in the shell and $\bar n$ is the mean number density of tracer haloes in the whole simulation box.
We realize that the mean value of the number overdensity, denoted as $\langle \Delta\rangle$, for a sample of targets, which are LGAs in this study, is essentially the cross-correlation function between the targets and the tracer haloes.
Furthermore, $\langle\Delta\rangle$ is proportional to the large-scale bias of the targets, assuming linear halo bias.

To prove the above two statements, we start by denoting the target sample as `T', the tracer halo sample as `t', and the dark matter density field as `m'.
Then, we recall that the cross-correlation function estimator in \citet{davisSurveyGalaxyRedshifts1983} is
\begin{equation}
  \xi_{\rm Tt} = \frac{N_{\rm Tt}}{N_{\rm Tr}} - 1
  \label{eq:ccf_estimator}
\end{equation}
where $N_{\rm Tt}$ and $N_{\rm Tr}$ are the number of T-t and T-r pairs, in which `r' represents the random points that have the same number density as the tracer haloes in the simulation box.
Here we fix the radial bin of the cross-correlation function to cover the range $5-15\hmpc$, so that we can define the cross-correlation function as
\begin{equation}
  \xi_{\rm Tt} = \frac{\langle n \rangle V}{\bar n V} - 1 = \frac{\langle n\rangle - \bar n}{\bar n} = \langle \Delta \rangle
\end{equation}
where $\langle n\rangle$ is the mean number density of tracer haloes around the targets, as in equation~\eqref{eq:delta}, and $V$ is the volume of a spherical shell spanning $5-15\hmpc$.

Then, the linear halo bias assumption gives
\begin{equation}
  \xi_{\rm Tt} = b_{\rm t}\xi_{\rm Tm},~~~{\rm where}~b_{\rm t}\equiv \xi_{\rm tm}/\xi_{\rm mm},
  \label{eq:b_def}
\end{equation}
where $\xi_{\rm Tt}$, $\xi_{\rm tm}$, and $\xi_{\rm Tm}$ are the cross-correlation function of target-tracer, tracer-matter, and target-matter, respectively. Similarly, the bias of the target sample is defined as $b_{\rm T}\equiv \xi_{\rm Tm}/\xi_{\rm mm}$.
Consequently, the ratio between $b_{\rm T}$ and $\langle \Delta\rangle$ is
\begin{equation}
  f \equiv {b_{\rm T}\over \langle \Delta\rangle} = {b_{\rm T} \over \xi_{\rm Tt}} = {\xi_{\rm Tm} / \xi_{\rm mm} \over b_{\rm t}\xi_{\rm Tm}} = {1\over b_{\rm t}\xi_{\rm mm}}
\end{equation}
Here $\xi_{\rm mm}$ is a constant for a fixed radial range.

Motivated by the linear relationship between halo bias $b$ and $\langle \Delta
  \rangle$, we define the large-scale bias $\tilde b \equiv f\Delta$ for each target, and the coefficient $f$ is calibrated to recover the nominal dependence of halo bias on halo mass, as demonstrated in Fig.~\ref{fig:figure/halo_bias}.
Setting $f=4.53$ recovers the halo bias relation of \citet{tinkerLargescaleBiasDark2010} to an accuracy better than 5\% over two orders of magnitude of halo mass, providing further support to the proof outlined earlier.
We note the difference that $\tilde b$ is defined for each individual target, while $b$ is defined for a sample of targets and estimated from its two-point correlation function (see equation~\eqref{eq:b_def}).
These two quantities are related through $b = \langle \tilde b \rangle$ over a sample of targets.

\subsubsection{Large-scale environment in redshift space}
\label{sec:large_scale_environment_in_redshift_space} 

In real observations, the measured position of a galaxy group, which is used as a tracer of the dark matter halo, is contaminated by the redshift space distortion effect, which biases the overdensity estimation.
Nevertheless, we find that a constant shift in the estimated large-scale LGA bias can eliminate this observational systematic.

In the redshift space, the number overdensity of tracer haloes is
\begin{equation}
  \Delta^{\rm RS} \equiv {n^{\rm RS} - \bar n\over \bar n}
  \label{eq:delta_rs}
\end{equation}
where $n^{\rm RS}$ is the number density of tracer haloes within $5-15\hmpc$ as measured in the redshift space.
We find that setting the large-scale bias to
$\tilde b = f(\Delta^{\rm RS} - c)$ with $c = 0.15$ recovers not only the halo bias relation in \citet{tinkerLargescaleBiasDark2010}, but exhibits an unbiased one-to-one relation to $f\Delta$ in different halo mass bins, as shown in Fig.~\ref{fig:figure/halo_bias}.
We note that this redshift space distortion correction is significant and cannot be simply ignored even though we are focusing on a quite large scale up to $15\hmpc$.
This correction shifts the halo bias by $cf\approx 0.68$, which corresponds to $\approx 85\%$ to $\approx 38\%$ of the halo bias relation in the mass range of $10^{11.5-13.5}\hMsun$.

The large-scale environment of the our LG is mapped by the 2MASS Redshift Survey \citep[2MRS,][]{skrutskieTwoMicronAll2006} and the galaxy group catalog \citep{limGalaxyGroupsLowredshift2017}.
The latter contains 59 galaxy groups with estimated halo masses above $10^{11.6}\hMsun$ within $5-15\hmpc$ in the redshift space around our MW.
After accounting for sky incompleteness due to the galactic extinction ($|b| < 10^\circ$), the effective tracer halo number is $\approx 74.3\pm{8.1}$ (see Appendix~\ref{sec:uncertainty_mask}) and the number overdensity is $\Delta_{\rm LG}\approx -0.44\pm {0.06}$, given that the mean tracer halo number density is $\bar n = 9.7\times 10^{-3}h^3{\rm Mpc}^{-3}$ in the ABACUSSUMMIT simulation.
Finally, we estimate the large-scale bias of our LG to be $\tilde b_{\rm LG}\approx -2.7\pm{0.3}$, placing it at the 1.4\% quantile of the LGA distribution.

\section{Relating local group analogues to the cosmic web}%
\label{sec:relating_local_group_analogs_to_cosmic_web}

With our sample of LGAs and their large-scale environment established, we can investigate the statistical relationship between them.
We first examine the spatial distribution of LGAs in cosmic web in \S\,\ref{sub:the_spatial_distribution_of_lga_in_cosmic_web}.
Then, we quantify the relationship between the coupling energy and the large-scale overdensity in \S\,\ref{sub:coupling_energy_and_large_scale_bias}, and the relationship between the LGA configuration and the anisotropic cosmic web in \S\,\ref{sub:coupling_energy_and_large_scale_anisotropy}.
Finally, \S\,\ref{sec:relative_velocity_and_large_scale_environment} presents the detailed modeling of the relative velocity distribution for the radial and the tangential components, respectively.

\subsection{The spatial distribution of LGA in cosmic web}
\label{sub:the_spatial_distribution_of_lga_in_cosmic_web} 

\begin{figure*}
  \begin{center}
    \includegraphics[width=0.95\linewidth]{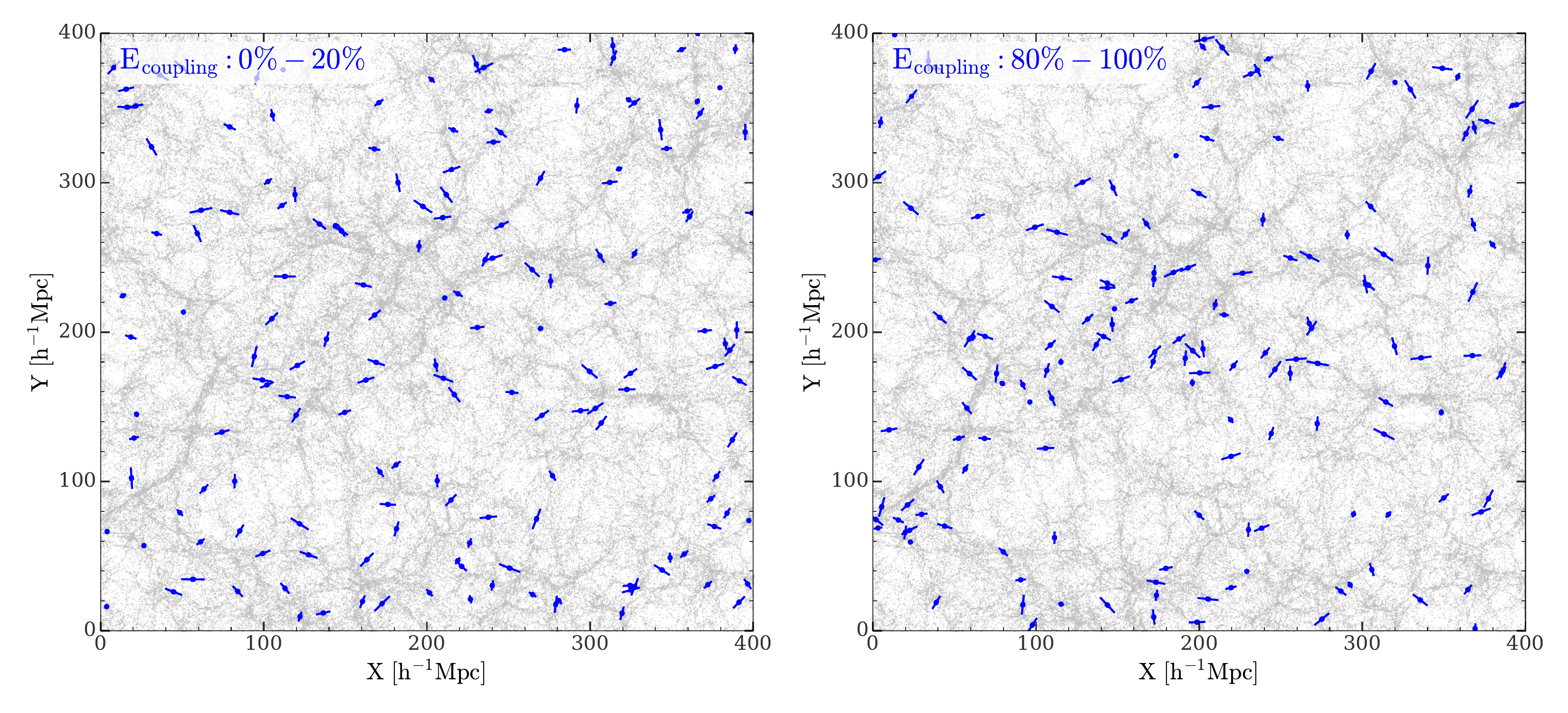}
  \end{center}
  \caption{
    The spatial distribution of LGAs in the cosmic web for the lowest 20\% (left) and highest 20\% (right) of $E_{\rm coupling}$, projected on the $X-Y$ panel over a depth of $10h^{-1}\rm Mpc$ along the $Z$-axis.
    In each panel, the blue line segments indicate the orientation of the halo pair in each LGA, and the length proportional to their projected separation by a factor of 10.
    The background color map shows the distribution of all tracer haloes.
    LGAs are preferentially distributed in filamentary and cluster structures, while maintaining an alignment to the filaments, especially for those with low $E_{\rm coupling}$.
    Comparing these two panels, LGAs with low $E_{\rm coupling}$ are more sparsely distributed, and those with high $E_{\rm coupling}$ exhibit stronger clustering.
  }
  \label{fig:lga_spatial_demo}
\end{figure*}

Fig.~\ref{fig:lga_spatial_demo} shows a slice of $400\hmpc\times 400\hmpc$ projected cosmic density field (thickness $10\,h^{-1}\rm Mpc$, shown in grey) in the background grey color.
The foreground blue line segments show the distribution of LGAs with the lowest 20\% $E_{\rm coupling}$ (left panel) and the highest 20\% $E_{\rm coupling}$ (right panel), and their orientation indicates the projected displacement vector connecting the two haloes of LGAs.
LGAs generally trace the matter distribution: most of the LGAs are in filamentary and cluster structures, while a few of them are in void regions.
In addition, comparing LGAs with different coupling energy, we find that LGAs with low-$E_{\rm coupling}$ are more sparsely distributed with a preference to filaments, while their counterparts with high-$E_{\rm coupling}$ exhibits a stronger spatial clustering strength, preferentially in cosmic nodes.
We also find that halo pairs in LGAs tend to align with the surrounding large-scale filaments.
In the following subsections, we will quantify the connection between LGAs and their large-scale environment.
These observations motivate more detailed quantification of the relationship between LGAs and their surrounding large-scale cosmic web.

\subsection{Coupling energy and large-scale environment}
\label{sub:coupling_energy_and_large_scale_bias} 

\begin{figure}
  \centering
  \includegraphics[width=1.0\linewidth]{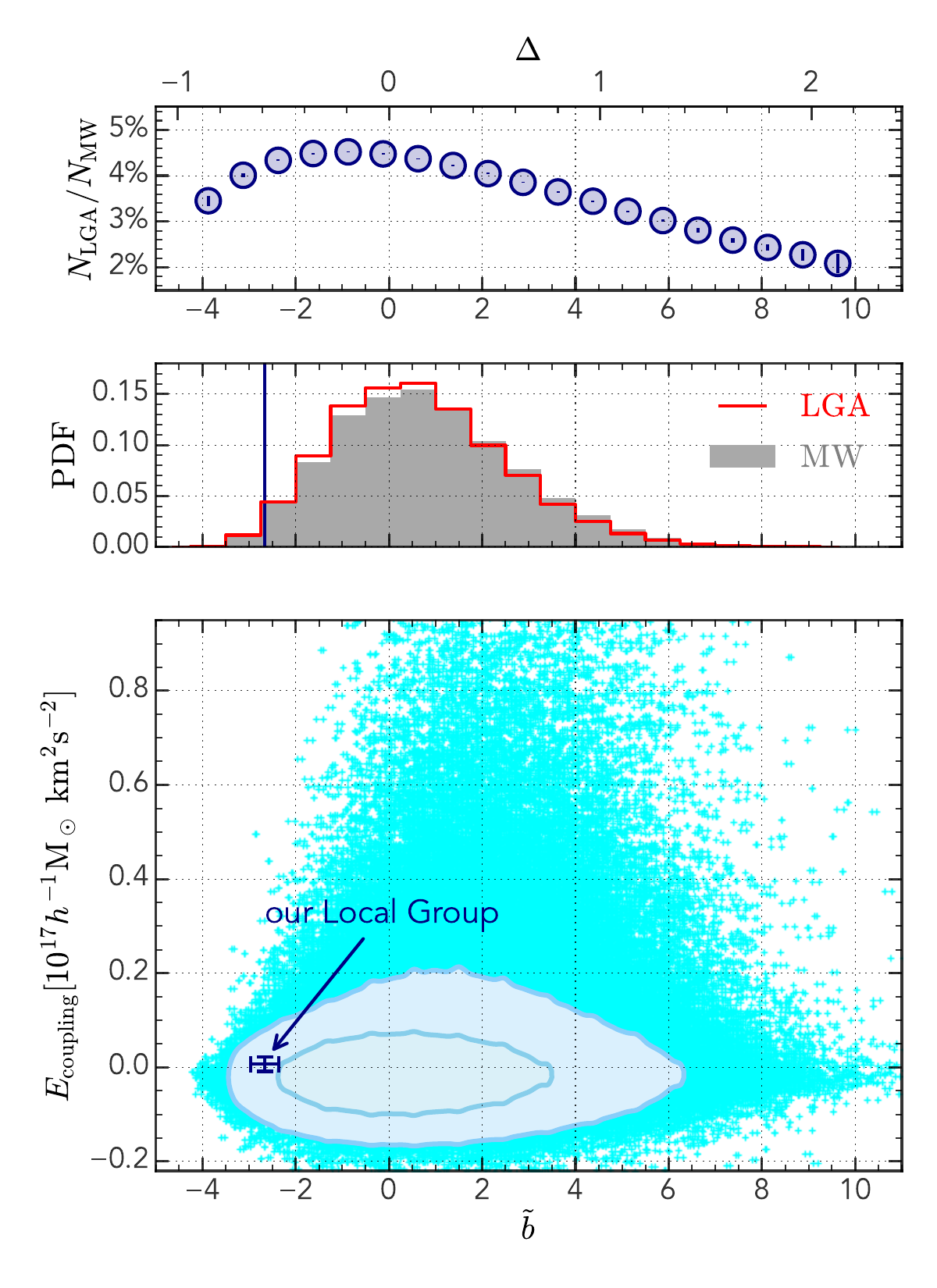}
  \caption{
    \textbf{Top panel:} The fraction of LGAs from all MW-mass haloes as a function of large-scale bias with error bar estimated from 25 ABACUSSUMMIT simulation boxes.
    \textbf{Middle panel:} The normalized distribution of the large-scale bias of MW-mass haloes in grey-filled histogram and of LGAs in red histogram.
    The navy vertical line shows the location of our Local Group.
    \textbf{Bottom panel:} The joint distribution of the LGA coupling energy, $E_{\rm coupling}$, and the large-scale LGA bias, $\tilde b$.
    The contour lines encompass 68 and 95 per cent of all LGAs, and the remaining 2 per cent LGAs are shown with cyan crosses.
    The location of our LG is shown with the large navy symbol.
    Overall, the fraction of LGA shows a moderate dependence on the large-scale environment, with a preference for under-dense regions.
    Compared to LGAs identified in simulation, our LG exhibits typical $E_{\rm coupling}$ and occupies a relatively under-dense environment.
  }%
  \label{fig:figure/eb_env_joint_distribution}
\end{figure}

\begin{figure}
  \centering
  \includegraphics[width=1.00\linewidth]{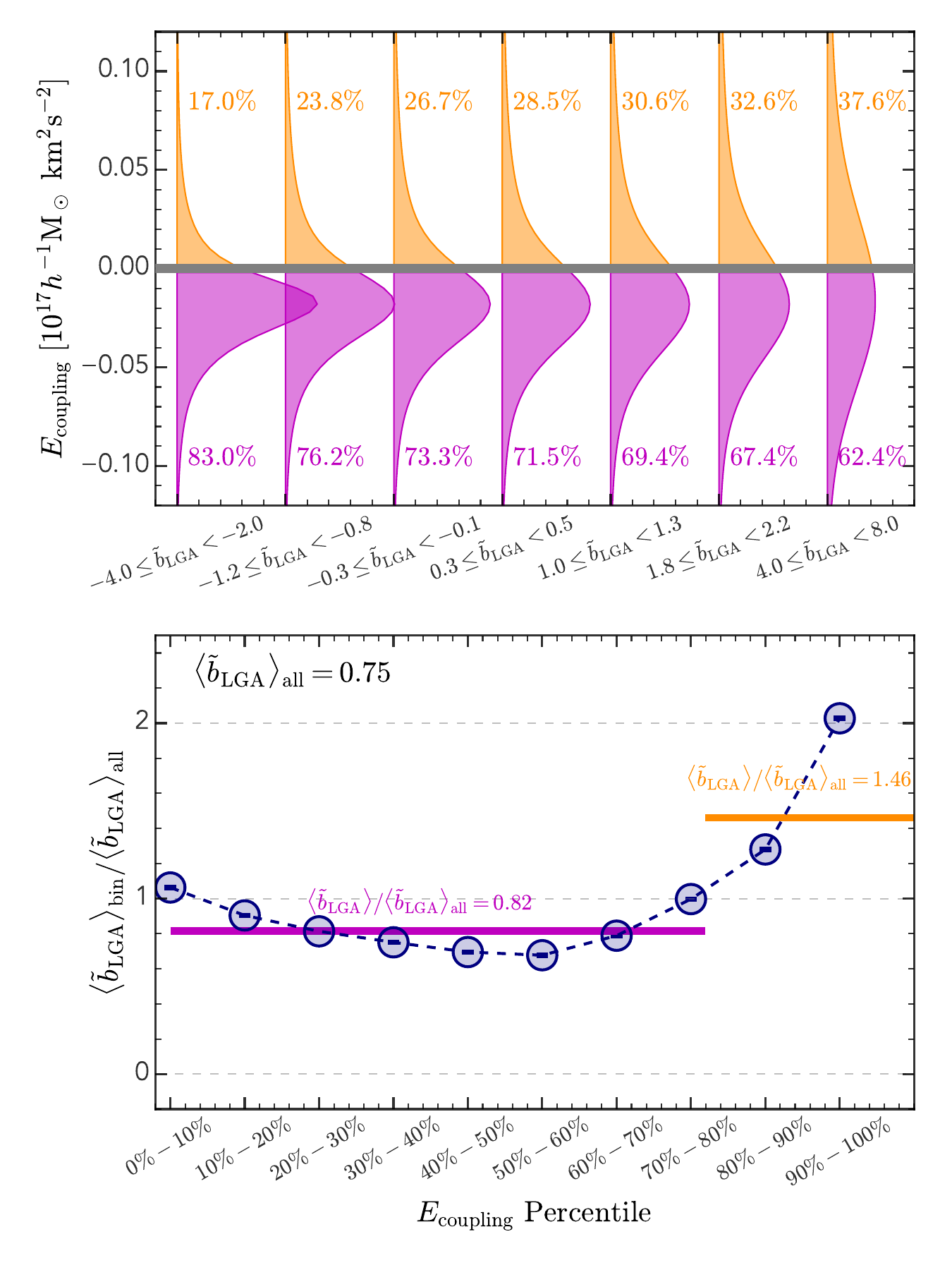}
  \caption{
    {\bf Top panel:} The normalized distribution of $E_{\rm coupling}$ in different large-scale bias bins.
    The dark orange and magenta colors are for $E_{\rm coupling} > 0$ and $E_{\rm coupling} < 0$, respectively, together with their relative fractions labeled. {\bf Bottom panel:} The mean value of large-scale bias in different $E_{\rm coupling}$ percentile bins normalized by the mean bias for all LGAs.
    The error bar is estimated with 25 ABACUSSUMMIT simulation boxes.
    The magenta and dark orange horizontal lines show the mean normalized bias for coupled and uncoupled LGAs, respectively.
    Overall, LGAs in over-dense environments exhibit typical $E_{\rm coupling}$ than those in under-dense regions.
    The mean bias for uncoupled LGAs exceeds that of coupled LGAs by approximately $1.46 / 0.82 - 1\approx 78\%$.
  }
  \label{fig:figure/relation_eb_env}
\end{figure}

The bottom panel of Fig.~\ref{fig:figure/eb_env_joint_distribution} presents the joint distribution of the coupling energy, $E_{\rm coupling}$, and the large-scale bias, $\tilde b$, with contour lines enclosing 68 and 95 per cent of LGAs.
The marginalized distribution of $\tilde b$ is shown as a navy histogram in the middle panel of Fig.~\ref{fig:figure/eb_env_joint_distribution}.
It spans the range from $\tilde b\approx -4$, which is the lowest-density region that can be probed by our tracer here, to $\tilde b\gtrsim 10$, which is comparable of the typical bias for galaxy clusters \citep{jingDependenceDarkHalo2007}.
Comparison with the $\tilde b$ distribution of all MW-mass haloes regardless of their environment (grey shaded histogram) reveals that LGAs (red histogram) prefer to live in marginally lower-density regions compared to MW-mass haloes.
This trend is elucidated further in the top panel of Fig.~\ref{fig:figure/eb_env_joint_distribution}, showing that LGAs prefer to reside in environments with $\tilde b\approx -1$ and declines towards both extremes.
This trend indicates two competing factors: low-density regions have fewer potential halo pairs, and high-density regions have higher chance to break the isolation criterion.
Finally, although the Local Group has a typical $E_{\rm coupling}$ compared to analogue systems, it resides in a low-density environment with $\tilde b\approx -2.7$, which is lower than 98.6 per cent of LGAs.

The correlation between the coupling energy and the large-scale overdensity is obscured by the large scatter in Fig.~\ref{fig:figure/eb_env_joint_distribution}.
To shed more light on this relationship, we present in the top panel of Fig.~\ref{fig:figure/relation_eb_env} the normalized distribution of $E_{\rm coupling}$ in bins of large-scale bias, $\tilde b$.
The uncoupled ($E_{\rm coupling} > 0$) and coupled ($E_{\rm coupling} < 0$) LGAs are colored differently with their fractions labeled.
The fraction of coupled LGAs ($\approx 83.0\%$) is the highest in the lowest-density bin and drops as $\tilde b$ increases to $\approx 62.4\%$ in the highest-density bin.

The bottom panel of Fig.~\ref{fig:figure/relation_eb_env} shows the mean large-scale LGA bias relative to $\langle \tilde b_{\rm LGA}\rangle_{\rm all}$ as a function of $E_{\rm coupling}$ percentiles.
Firstly, the mean individual bias for all LGAs, denoted as $\langle \tilde b_{\rm LGA}\rangle_{\rm all}$, is approximately $0.75$.
This is slightly lower than that of all MW-haloes ($\approx 0.82$; see Fig.~\ref{fig:figure/halo_bias}).
This difference is expected since LGAs prefer to live in slightly lower-density regions compared to the MW-mass haloes (see Fig.~\ref{fig:figure/eb_env_joint_distribution}).
Secondly, the mean large-scale bias of LGAs depends on the $E_{\rm coupling}$ of LGAs.
The LGAs with the lowest-10\% $E_{\rm coupling}$ have a mean large-scale bias similar to all LGAs combined, $\langle \tilde b_{\rm LGA}\rangle_{\rm all}$.
The mean large-scale LGA bias declines to $\approx 0.6\langle \tilde b_{\rm LGA}\rangle_{\rm all}$ as $E_{\rm coupling}$ increases to its median value.
Then, it increases rapidly to $\approx 2.0\langle \tilde b_{\rm LGA}\rangle_{\rm all}$, reached at the top-$10\% E_{\rm coupling}$ values.
Overall, the mean individual bias of uncoupled LGAs ($E_{\rm coupling}
  < 0$) is about $1.46\langle \tilde b_{\rm LGA}\rangle_{\rm all}$, and the value for coupled LGAs ($E_{\rm coupling} > 0$) is about $0.82\langle \tilde b_{\rm LGA}\rangle_{\rm all}$.
Their biases differ by a factor of $\approx 1.78$.
This difference is significant as the halo assembly bias (i.e. the bias difference of haloes with different formation time) is only $\approx 1.60$ between haloes in the top-20\% and bottom-20\% highest formation redshift in this mass range \citep[see][]{gaoAgeDependenceHalo2005}.

\subsection{Coupling energy and large-scale anisotropy}
\label{sub:coupling_energy_and_large_scale_anisotropy} 

\begin{figure}
  \begin{center}
    \includegraphics[width=0.45\textwidth]{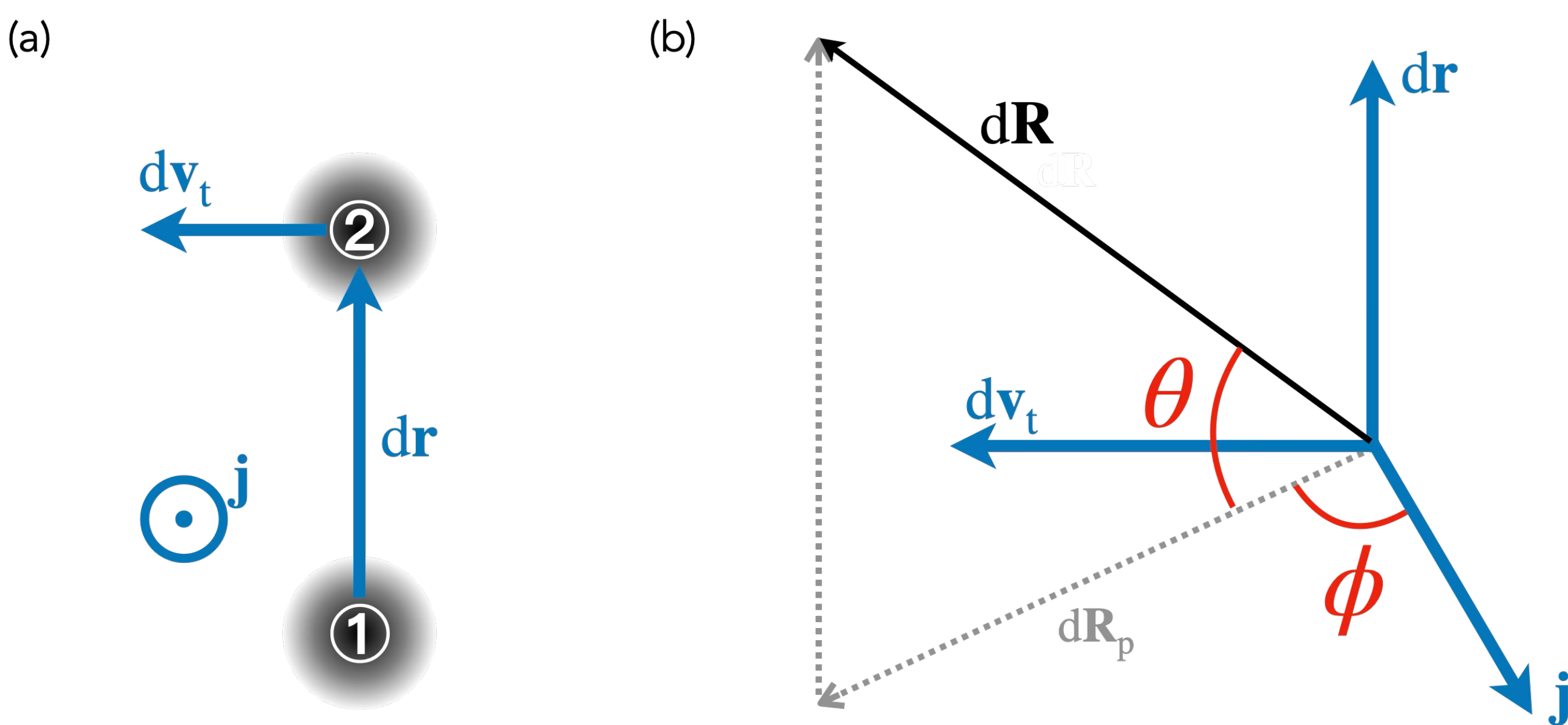}
  \end{center}
  \caption{
  Illustration of the coordinate system used for the LGA anisotropy analysis.
  \textbf{(a):} $\mathrm d\mathbf r$ is the displacement vector from the primary halo to the secondary halo, and $\mathrm d\mathbf v_{\rm t}$ is the tangential velocity of the secondary halo relative to the primary halo.
  The specific orbital angular momentum $\mathbf j\equiv \mathrm d\mathbf r\times \mathrm d \mathbf v_{\rm t}$ points outward, perpendicular to the paper.
  \textbf{(b):} A Cartesian coordinate system defined by $\mathrm d\mathbf r$, $\mathrm d\mathbf v_{\rm t}$ and $\mathbf j$.
  For a tracer halo with position $\mathrm d\mathbf R$ relative to the primary halo, two angles to characterize its angular position are calculated: $\theta~\in~[-90^\circ,~+90^\circ]$ and $\phi~\in~(-180^\circ,~+180^\circ]$.
  }
  \label{fig:demo_coordinate}
\end{figure}

\begin{figure*}
  \begin{center}
    \includegraphics[width=1.0\textwidth]{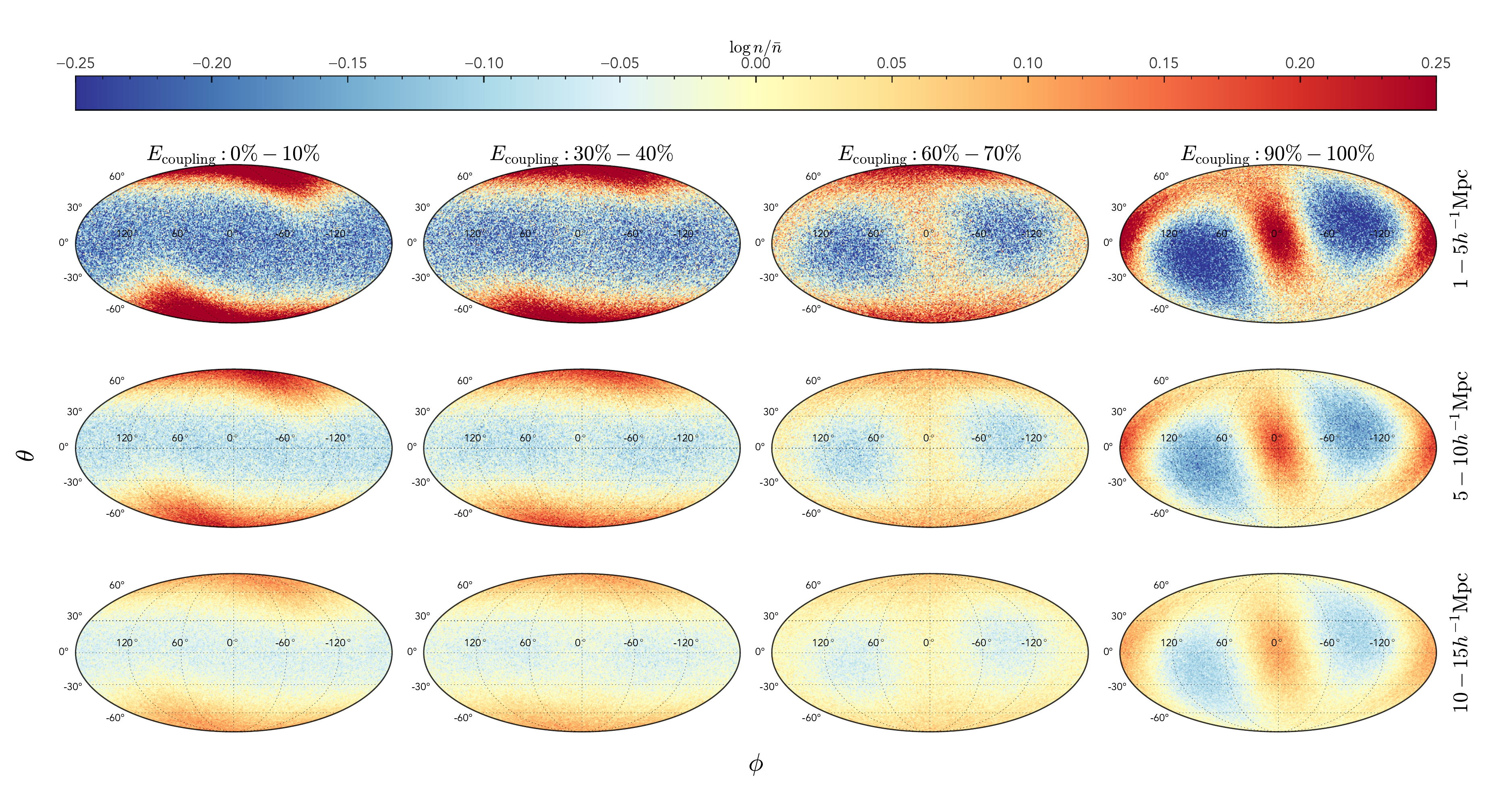}
  \end{center}
  \caption{
    The deviation from the mean number density distribution of tracer haloes on the Mollweide projection as a function of the polar angle, $\theta$, and the azimuthal angle, $\phi$.
    The coordinate is built so that the displacement vector points to $\theta =90^\circ$, and the relative velocity of the secondary halo to the halo in question points to $(\theta, \phi) = (0^\circ, 90^\circ)$, and the specific orbital angular momentum points to $(\theta, \phi) = (0^\circ, 0^\circ)$.
    The color encodes the deviation from the mean tracer halo number density on the logarithmic scale, $\log n/\bar n$, ranging from -0.25 dex to 0.25 dex (as indicated by the colour bar).
    LGAs are binned according to the coupling energy $E_{\rm coupling}$ (from left to right), and results for three difference radial shells ($1-5\hmpc$, $5-10\hmpc$, and $10-15\hmpc$) are shown from top to bottom respectively.
    This figure demonstrates that, for a fixed radial shell, the anisotropic pattern transforms from strong polar anisotropy in the lowest-$E_{\rm coupling}$ bin to strong azimuthal anisotropy in the highest-$E_{\rm coupling}$ bin.
  }
  \label{fig:healpix_logn}
\end{figure*}

\begin{figure*}
  \begin{center}
    \includegraphics[width=1.0\linewidth]{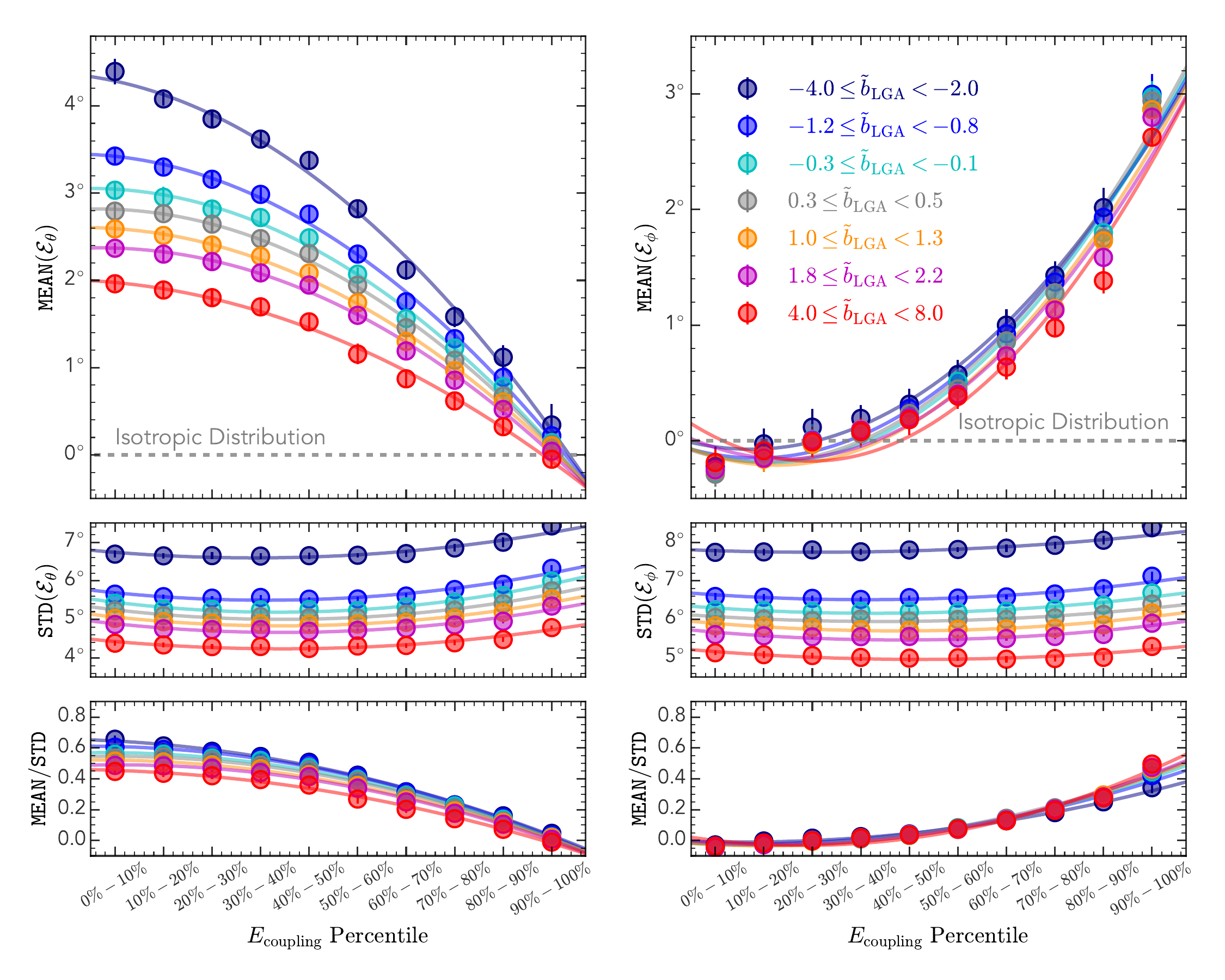}
  \end{center}
  \caption{
    \textbf{Top panels:} The polar anisotropy, $\mathcal E_\theta$ (left), and the azimuthal anisotropy, $\mathcal E_\phi$ (right), measured using tracer haloes within $5-15~\hmpc$, shown as a function of the $E_{\rm coupling}$ percentiles, in different large-scale overdensity bins.
    The horizontal grey dashed line corresponds to the expectation for an isotropic distribution.
    \textbf{Middle panels:} The standard deviation of $\mathcal E_\theta$ and $\mathcal E_\phi$ across individual LGAs. \textbf{Bottom panels:} The ratio between the mean value and the standard deviation of $\mathcal E_\theta$ and $\mathcal E_\phi$, quantifying the average signal-to-noise ratio for detecting these anisotropies from a single system.
    Error bars in all panels are estimated from 25 ABACUSSUMMIT simulation boxes.
    Solid curves represent second-order polynomial fits to the data points of the corresponding colour for visual guidance.
    Overall, the polar anisotropy is strongest for LGAs with the lowest $E_{\rm coupling}$, decreasing steadily and becoming consistent with isotropy in the highest $E_{\rm coupling}$ bin.
    It also shows moderate dependence on environment, being stronger in under-dense regions.
    In contrast, the azimuthal anisotropy peaks in the highest $E_{\rm coupling}$ bin, declines toward lower $E_{\rm coupling}$, and exhibits negligible environmental dependence.
  }
  \label{fig:e_theta_phi}
\end{figure*}

As we have established the connection between LGAs and their large-scale overdensity, we proceed to investigate the relationship to the anisotropic part of the cosmic web.
There are many different methods to relate the target of interest to the anisotropic cosmic web.
One way is to identify the cosmic web first, and then study its relationship to the target of interest \citep[e.g.][]{patiri2006alignment, codis2012connecting}.
However, this method is subject to the specific algorithm that is used to identify the cosmic web \citep[see][for a review]{libeskindTracingCosmicWeb2018}.
The other kind of methods is to start from identifying some intrinsic vectors from the target of interest, and then measure the anisotropic distribution of large-scale tracers with respect to these vectors.
This is exemplified by \citet{faltenbacherSpatialKinematicAlignments2008}, in which they found that satellite substructures prefer to align to the major axis of the central substructure, and this correlation extends to six times the virial radius \citep[see][for similar studies]{zhangSpinOrientationDark2009, liDetectionLargescaleAlignment2013}.

Here we quantify the connection between LGAs and the anisotropic part of cosmic web in three steps.
Firstly, we identify two intrinsic vectors for each LGA system as demonstrated in the left diagram of Fig.~\ref{fig:demo_coordinate}: the displacement vector, $\mathrm d\mathbf r$, from the primary halo to the secondary halo, and the tangential velocity of the secondary halo with respect to the primary halo, $\mathrm d\mathbf v_{\rm t}$.
The cross product of these two vectors produces the specific orbital angular momentum of the whole system.
These three vectors form the principle axes of the Cartesian coordinate system on which we map the surrounding large-scale tracer haloes.

The angular position of any tracer halo, which relative position to the halo in question is $\rmd\mathbf R$, can be characterised by two angles, as shown in the right diagram of Fig.~\ref{fig:demo_coordinate}: the polar angle $\theta$ that ranges from $-90^\circ$ to $+90^\circ$ and the azimuthal angle $\phi$ that ranges from $-180^\circ$ to $180^\circ$.
They are calculated as
\begin{align}
  \theta & =\arcsin\left({\mathrm d\mathbf r\cdot \mathrm d\mathbf R\over\|\mathrm d\mathbf r\|\cdot\|\mathrm d\mathbf R\|}\right) \\
  \phi   & =
  \begin{cases}
    \phi^\prime  & {\rm ~~if~~} \mathrm d\mathbf R_{\rm p}\cdot \mathrm d\mathbf v_{\rm t} \geq 0 \\
    -\phi^\prime & {\rm ~~if~~} \mathrm d\mathbf R_{\rm p}\cdot \mathrm d\mathbf v_{\rm t} < 0,   \\
  \end{cases}
\end{align}
in which
\begin{equation}
  \phi^\prime = \arccos\left({\mathrm d\mathbf R_{\rm p}\cdot \mathbf j\over \|\mathrm d\mathbf R_{\rm p}\|\cdot\|\mathbf j\|}\right),\\
\end{equation}
where $\mathrm d\mathbf R_{\rm p}  =\mathrm d\mathbf R- {\left(\mathrm d\mathbf R\cdot \mathrm d\mathbf r\right)\mathrm d\mathbf r/ \|\mathrm d\mathbf r\|^2}$. Within this definition, $\theta=90^\circ$ is the direction from the primary halo to the secondary halo, and $\theta=-90^\circ$ is the opposite direction. $\phi=0^\circ$ is the direction of the orbital spin.

Fig.~\ref{fig:healpix_logn} presents the deviation from the mean tracer halo number density, $\log (n/\bar n)$, as function of $(\theta,\phi)$, split by four $E_{\rm coupling}$ coupling percentiles (left to right) and three different radial shells centered on the primary halo in the LGA system (from top to bottom).
We use the HEALPix\footnote{http://healpix.sf.net} package with the Mollweide projection \citep{Zonca2019, 2005ApJ...622..759G}.
The colour mapping encodes the deviation from the mean tracer halo number density, $\log (n/\bar n)$, where blue is for under-dense regions and red is for over-dense regions.

Firstly, the anisotropic pattern is similar in different radial shells, with the amplitude of the anisotropy decreasing as the distance to the radial shell increases.
This means that the correlation between the LGA velocity structure and the surrounding large-scale structure is weaker at greater radial distance, as expected.
Secondly, the anisotropic pattern changes with the coupling energy (left to right panels in Fig.~\ref{fig:healpix_logn}).
Low-$E_{\rm coupling}$ LGAs have strong polar anisotropy, for which tracer haloes are preferentially aligned with the displacement vector of the halo pair.
Meanwhile, high-$E_{\rm coupling}$ LGAs have strong azimuthal anisotropy, for which tracer haloes are preferentially located on a plane defined by the displacement vector and the specific orbital angular momentum vector.
The change from strong polar anisotropy to strong azimuthal anisotropy is progressive, as one can see from the intermediate $E_{\rm coupling}$ percentiles (middle panels of Fig.~\ref{fig:healpix_logn}).

The maps in Fig.~\ref{fig:healpix_logn} effectively illustrate the anisotropic pattern, but a precise quantification of the anisotropy requires some more detailed statistical analysis.
We assess the degree of anisotropy by integrating the $\log (n/\bar n)$ distribution in the  $(\theta, \phi)$ space to get the marginalized distributions for $\theta$ and $\phi$ individually, and comparing their statistics to the expected values from the isotropic distribution.
To this end, we define two statistics to quantify the polar anisotropy and the azimuthal anisotropy, respectively, which are
\begin{align}
  \mathcal E_\theta & = \left\langle\lvert\theta\rvert\right\rangle_{\rm tracer} - \left\langle\lvert \theta\rvert\right\rangle_{\rm isotro}\label{eq:e_theta} \\ \mathcal E_\phi &=
  \left\langle\left\lvert\lvert\phi\rvert -\frac{\pi}{2}
  \right\rvert\right\rangle_{\rm tracer} -
  \left\langle\left\lvert\lvert\phi\rvert -\frac{\pi}{2}
  \right\rvert\right\rangle_{\rm isotro}\label{eq:e_phi}
\end{align}
where the bracket $\langle ...\rangle$ is defined differently for different subscripts, as in
\begin{align}
  \langle f\rangle_{\rm tracer} & \equiv \frac{1}{N_{\rm tracer}}\sum_{i=1}^{N_{\rm tracer}}f_i \\
  \langle f\rangle_{\rm isotro} & \equiv
  \int_{-\pi/2}^{\pi/2}\cos\theta\mathrm d\theta\int_{-\pi}^{\pi}\mathrm d\phi f \label{eq:expected_iso}
\end{align}

$\theta_i$ is the polar angle illustrated in the right diagram of Fig.~\ref{fig:demo_coordinate} for the $i$-th tracer halo. $\mathcal E_{\theta}$ quantifies the deviation from the isotropic distribution, where $\langle \lvert \theta\rvert\rangle_{\rm isotro}=\pi / 2 - 1$ is the expected value for the isotropic distribution.
Generally, $\mathcal E_\theta$ ranges from $1 - \pi / 2$ to $1$; $\mathcal E_\theta > 0$ indicates that tracer haloes tend to be aligned to the displacement vector, $\mathrm d\mathbf r$, as exemplified by the leftmost column in Fig.~\ref{fig:healpix_logn}.
Meanwhile, $\mathcal E_\theta < 0$ indicates that the tracer haloes tend to gather around the plane of $\theta=0$, perpendicular to $\mathrm d\mathbf r$.

Similarly, $\phi_i$ is the azimuthal angle illustrated in the right diagram of Fig.~\ref{fig:demo_coordinate}. $\mathcal E_\phi$ quantifies the deviation from the isotropic distribution, where $\langle \lvert\lvert \phi\rvert - \pi / 2\rvert\rangle_{\rm isotro} = \pi /4$ is the expected value for the isotropic distribution.
Generally, $\mathcal E_\phi \in [-\pi / 4,~\pi / 4]$: $\mathcal E_\phi > 0$ means that the tracer haloes tend to gather around the plane of $\phi=0^\circ/\pm 180\ddeg$, as seen in the rightmost column of Fig.~\ref{fig:healpix_logn}.
Meanwhile, $\mathcal E_\phi < 0$ corresponds to the opposite situation in which the tracer haloes tend to gather around the plane of $\phi=\pm 90\ddeg$.

Statistics of the inferred $\mathcal E_\theta$ and $\mathcal E_\phi$ are shown in Figs.~\ref{fig:e_theta_phi} (left and right panels respectively).
The top panels show the mean values of $\mathcal E_\theta$ and $\mathcal E_\phi$ of LGAs of different coupling energies split by large-scale overdensity bins.
The horizontal grey dashed line at zero is the expected value for the isotropic distribution, so that the deviation from the zero quantifies the degree of anisotropy.
The lower panels show the standard deviation of $\mathcal E_\theta$ and $\mathcal E_\phi$ and the ratio of the mean values to the standard deviations, which is indicative of the signal-to-noise ratio of detecting such anisotropic signals using a single LGA system.

The mean values of $\mathcal E_\theta$ and $\mathcal E_\phi$ depend on the coupling energy, $E_{\rm coupling}$, but in opposite ways.
First, the polar anisotropy quantified by $\mathcal E_\theta$ is the highest ($\mathcal E_\theta\approx 2.0\ddeg -4.4\ddeg$) at the lowest $E_{\rm coupling}$ bin, with a moderate dependence on the large-scale overdensity in a sense that lower density regions have stronger polar anisotropy.
Then, with increasing $E_{\rm coupling}$, the polar anisotropy declines until becoming consistent to the isotropic distribution ($\mathcal E_\theta \approx 0\ddeg$), regardless of the large-scale overdensity.
The azimuthal anisotropy quantified by $\mathcal E_\phi$ behaves in the opposite way.
It is consistent with the isotropic distribution ($\mathcal E_\phi = 0$) for the lowest $E_{\rm coupling}$ bins and progressively becomes larger for higher $E_{\rm coupling}$ percentile bins, reaching $\mathcal E_\phi\approx 3.0\ddeg$.
An additional difference is that the azimuthal anisotropy has only a minor dependence on the large-scale overdensity, unlike the polar anisotropy.

Despite that both the polar and azimuthal anisotropies are statistically significant for the lowest and highest $E_{\rm coupling}$ percentile bins, the variance, which is quantified by the standard deviation of $\mathcal E_\theta$ and $\mathcal E_\phi$, among individual LGA systems is also significant, as shown in the middle panels of Fig.~\ref{fig:e_theta_phi}.
In addition, the variances of both statistics depend on the large-scale overdensity, as the number density of tracer haloes is indicative of the sampling noise, with negligible dependence on $E_{\rm coupling}$.
Moreover, the variance is larger than the mean deviation from the isotropic distribution.
This means that it is challenging to detect such anisotropy from a single system using tracer haloes above $10^{11.6}\hMsun$.
Nevertheless, we can increase this signal-to-noise ratio by either using more tracer haloes, which will become available with deeper galaxy surveys, or stacking multiple LGAs together, which will be explored in a future paper.

Beyond the marginalized polar and azimuthal anisotropies, we also notice some asymmetric patterns in Fig.~\ref{fig:healpix_logn}.
In the lowest $E_{\rm coupling}$ bin, the overdensity region is enhanced at $(\theta \approx -90\ddeg,~\phi\approx 60\ddeg)$ and $(\theta\approx 90\ddeg,~\phi\approx -60\ddeg)$, creating a tilted overdensity pattern.
Conversely, the highest $E_{\rm coupling}$ bins show an opposite tilt orientation.
We speculate that these tilted patterns generate torques differentially that could affect LGA spins.
In low-$E_{\rm coupling}$ systems, the torques would act to decelerate the spin, maintaining alignment with large-scale structure and the polar anisotropy.
In high-$E_{\rm coupling}$ systems, the torques would instead accelerate the spins, potentially erasing the polar anisotropy over time.

\subsection{Relative velocity and large-scale environment}
\label{sec:relative_velocity_and_large_scale_environment} 

\begin{figure*}
  \centering
  \includegraphics[width=0.95\linewidth]{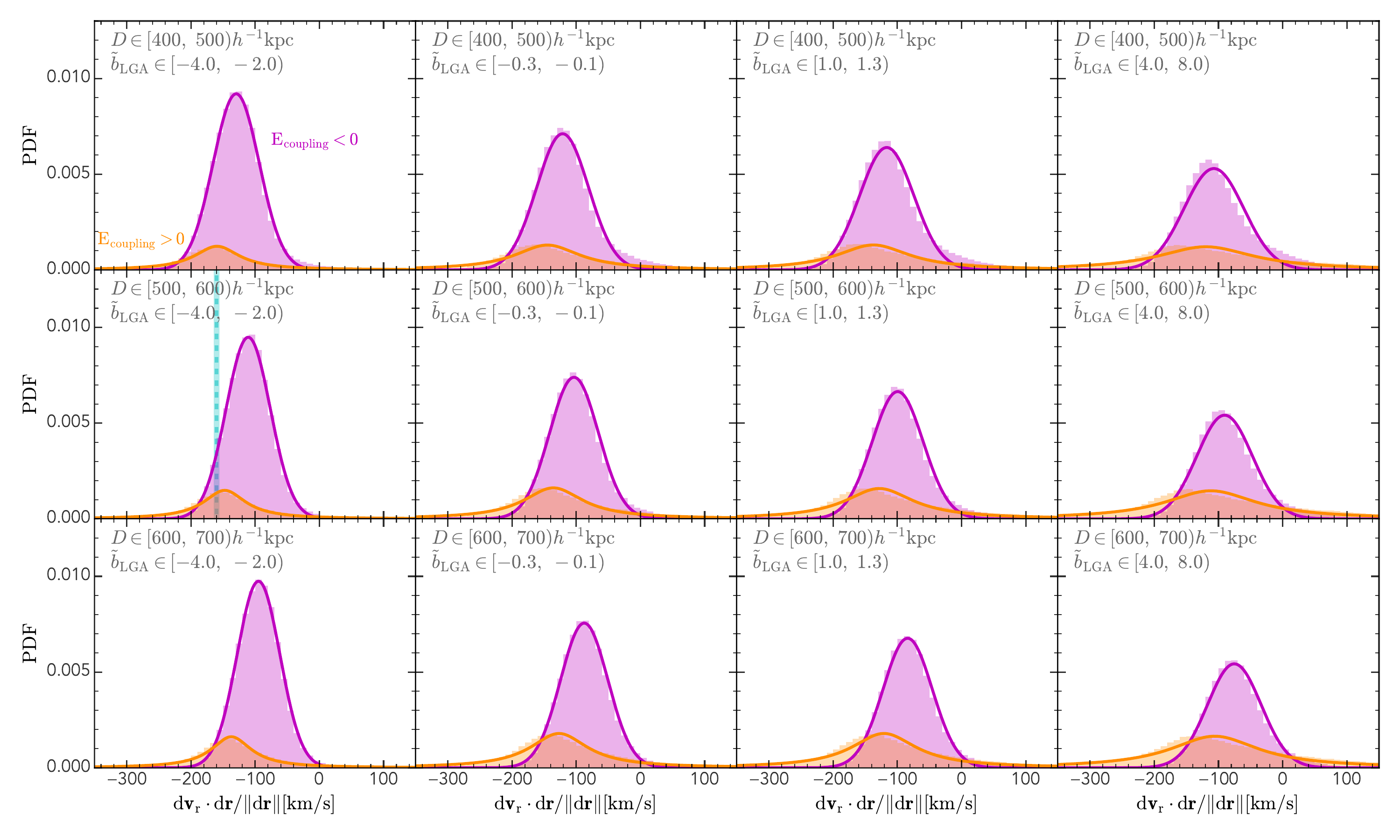}
  \caption{
    The normalized distributions of radial velocity between halo pairs in LGAs, shown for different separations (top to bottom) and large-scale bias bins (left to right).
    The magenta and orange histograms are for coupled and uncoupled LGAs, respectively.
    The solid lines are the best-fitting results with Student's t-distribution (see equation~\eqref{eq:student}), where the discrete parameter $\nu_{\rm r}$ is set to $\infty$ and 1 for coupled and uncoupled LGAs, respectively.
    The vertical dashed line in the middle-left panel shows the relative radial peculiar velocity of our LG, and the shaded region shows the uncertainty, which is too small to be distinguished from the width of the vertical dotted line.
  }%
  \label{fig:figure/radial_vel_fitting}
\end{figure*}

\begin{figure*}
  \begin{center}
    \includegraphics[width=0.95\linewidth]{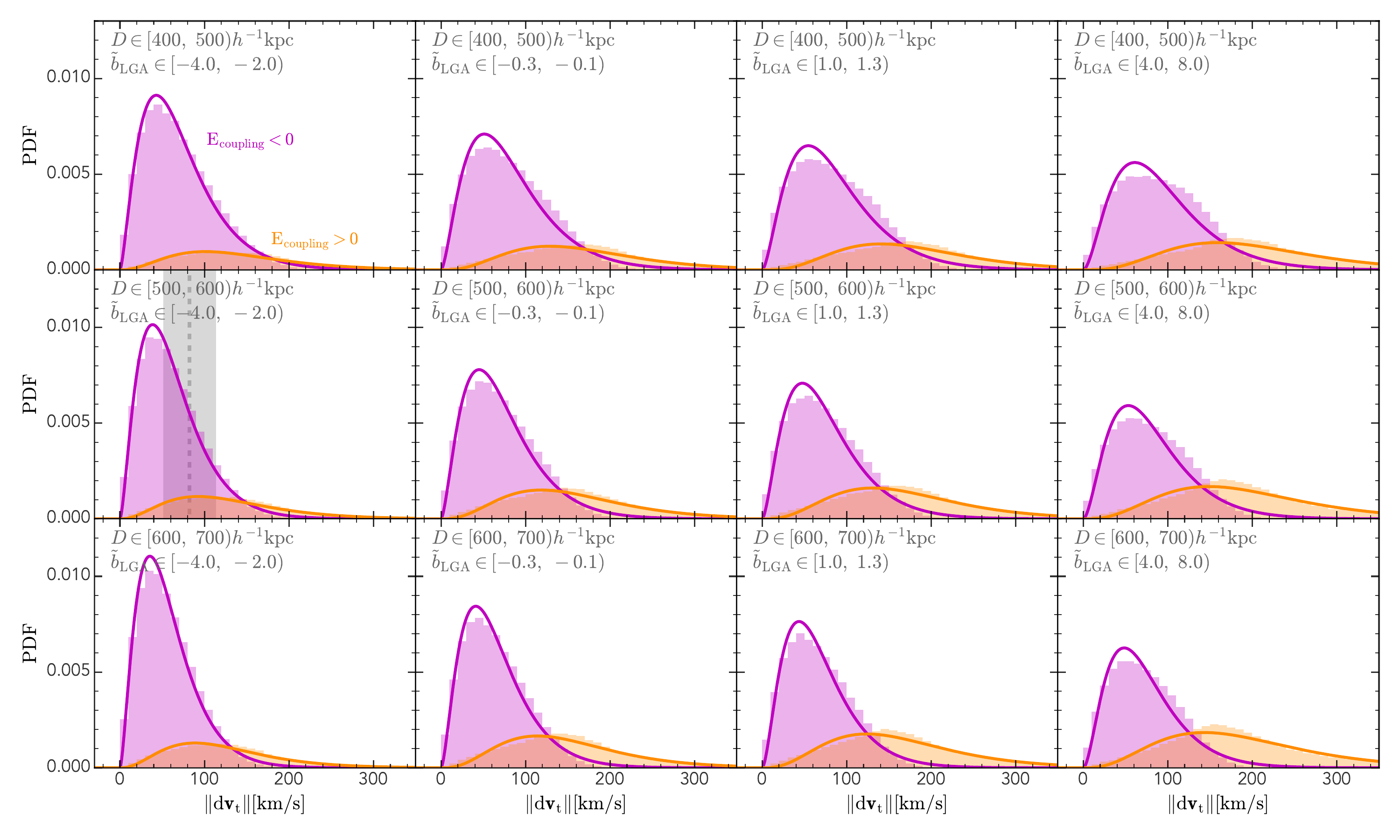}
  \end{center}
  \caption{
    Same as Fig.~\ref{fig:figure/radial_vel_fitting}, but for the tangential velocity.
    Both coupled ($E_{\rm coupling} > 0$) and uncoupled ($E_{\rm coupling} < 0$) LGAs are fitted with Gamma distributions (see equation~\eqref{eq:gamma}).
    The vertical dashed line and the shaded hatched region show the tangential velocity measurement of our Local Group, together with its uncertainty.
  }
  \label{fig:tangential_vel_total_fitting}
\end{figure*}

\begin{figure*}
  \begin{center}
    \includegraphics[width=0.95\linewidth]{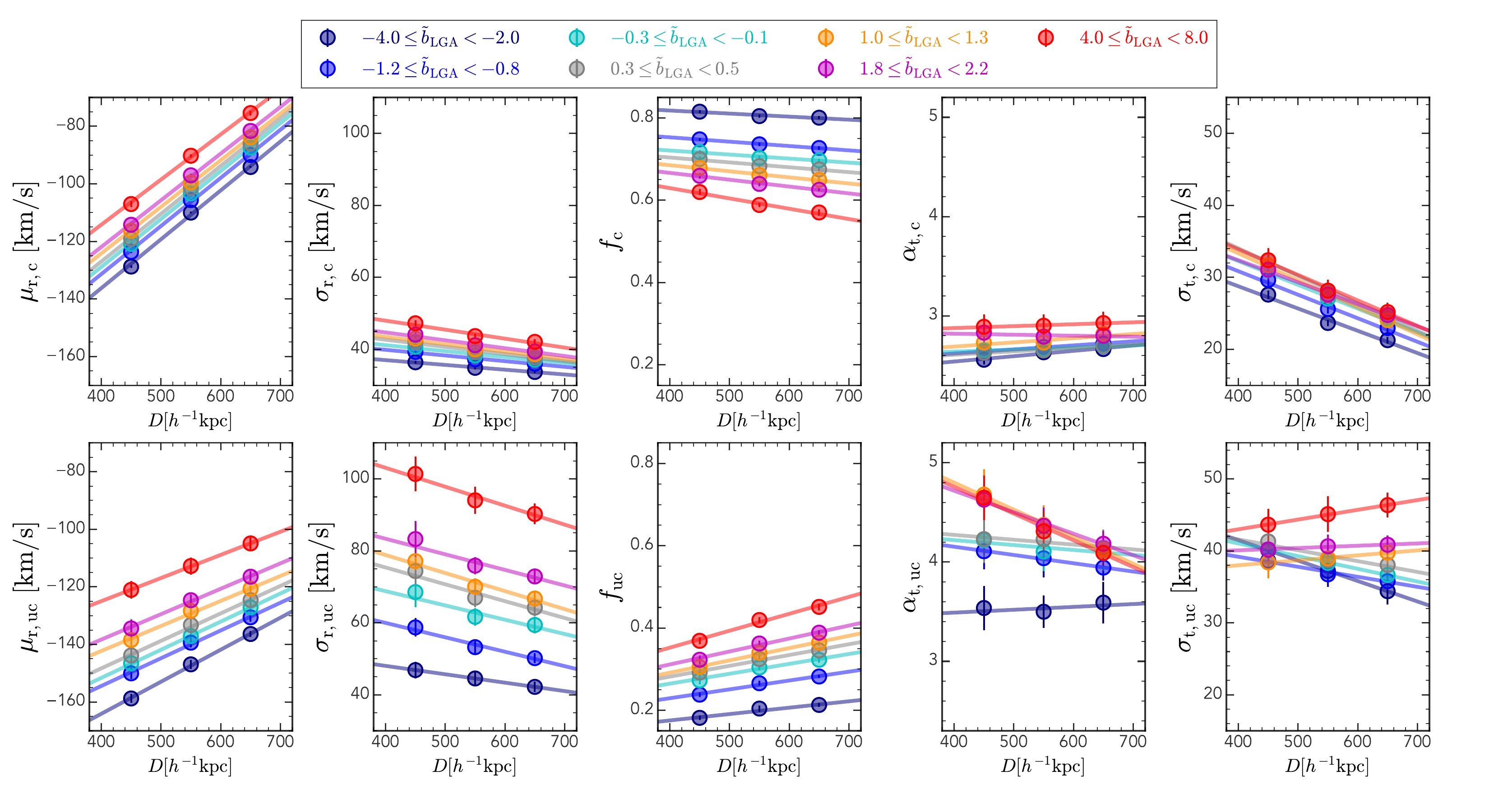}
  \end{center}
  \caption{
    The dependence of the fitting parameters in Figs.~\ref{fig:figure/radial_vel_fitting} and \ref{fig:tangential_vel_total_fitting} on the LGA separation and large-scale bias for coupled ($E_{\rm coupling} < 0$; top panels) and uncoupled ($E_{\rm coupling} > 0$; bottom panels) LGAs.
  }
  \label{fig:vel_dependence}
\end{figure*}

We have seen that the coupling energy, which reflects the total relative velocity of the halo pair, is strongly correlated to the large-scale environment.
The anisotropic nature of LG-like systems motivates us to investigate the radial and tangential components of the relative velocity (see equations~\ref{eq:vr} and \ref{eq:vt}) and their relationship with the large-scale environment.

Figs.~\ref{fig:figure/radial_vel_fitting} and~\ref{fig:tangential_vel_total_fitting} show the distribution of radial and tangential relative velocities, with results for coupled systems ($E_{\rm coupling} < 0$) and uncoupled systems ($E_{\rm coupling} > 0$) presented separately.
This is motivated by their markedly different distributions, as we will see later, and no single simple distribution function can adequately describe their combined behaviour.
We group the samples according to their separation and large-scale environment as we found these factors showed the strongest effect on the distributions.

We find that the radial velocity distributions can be described rather well by the student's $t$-distribution, which parametric form is
\begin{equation}
  p(x\mid \nu_{\rm r}, \mu_{\rm r}, \sigma_{\rm r}) = {1\over \sqrt{\nu_{\rm r}} ~ \mathrm B\left(1/2, \nu_{\rm r}/2\right) ~ \sigma_{\rm r}}\left[1 + \frac{1}{\nu_{\rm r}}\left(\frac{x-\mu_{\rm r}}{\sigma_{\rm r}}\right)\right]^{-{\nu_{\rm r} + 1 \over 2}}
  \label{eq:student}
\end{equation}
where $\nu_{\rm r}$ is the number of degrees of freedom, $\mathrm B$ is the beta function, $\mu_{\rm r}$ and $\sigma_{\rm r}$ quantifies the mean and variance of the distribution, and the subscript `r' denotes the radial velocity.
We found that setting $\nu_{\rm r}=\infty$, effectively reducing it to a Gaussian distribution, is required to match the coupled population ($E_{\rm coupling} < 0$), while the uncoupled population ($E_{\rm coupling} > 0$) requires $\nu_{\rm r}=1$, which is known as the Cauchy distribution, due to the presence of long broad tails.
Hereafter we use the subscript `c' and `uc' for the parameters to differentiate between these two situations.

The distribution of the tangential velocity, $\|\mathbf v_{\rm t}\|$, can be described by the Gamma distribution, which is
\begin{equation}
  p(x\mid \alpha_{\rm t},\sigma_{\rm t}) = {1\over
  \Gamma(\alpha_{\rm t})~\sigma_{\rm t}}\left({x\over \sigma_{\rm t}}\right)^{\alpha_{\rm t} - 1}
  \exp\left(-{x\over \sigma_{\rm t}}\right)
  \label{eq:gamma}
\end{equation}
where $\alpha_{\rm t}$ and $\sigma_{\rm t}$ are two parameters, and the mean and the variance of this distribution are $\alpha_{\rm t}\sigma_{\rm t}$ and $\alpha_{\rm t}\sigma_{\rm t}^2$, respectively.
The subscript `t' denotes the tangential velocity.

The fits to the distributions are shown with solid lines in Figs.~\ref{fig:figure/radial_vel_fitting} and \ref{fig:tangential_vel_total_fitting}.
The corresponding best-fit parameters are shown in Fig.~\ref{fig:vel_dependence} as a function of halo pair separation and large-scale environment.

The radial velocity distribution in Fig.~\ref{fig:figure/radial_vel_fitting} and the best-fitting parameters on the left four panels of Fig.~\ref{fig:vel_dependence} show several features.
Firstly, most LGAs have negative radial velocities of $\approx -100\kms$, which means that the two halo are approaching.
This is consistent with the timing mass argument \citep[see][]{kahn1959intergalactic, liMassesLocalGroup2008, fattahiAPOSTLEProjectLocal2016, strigariTimingMassLocal2025}.
The small fraction of halo pairs that are moving away from each other are predominantly found in the high-density region.
In addition, systems with the highest approaching velocity ($\approx -300\kms$) occur in the high-density regions.
Secondly, the uncoupled population ($E_{\rm coupling} > 0$) has a much larger dispersion than the coupled population ($E_{\rm coupling} < 0$).
Moreover, the dispersion parameter for uncoupled LGAs, $\sigma_{\rm uc}$, depends on the large-scale environment, where LGAs in dense regions have greater dispersion in their radial velocity distribution.
Finally, coupled LGAs have slightly lower approaching velocity compared to uncoupled LGAs and the mean approaching velocity is lower for halo pairs with larger separation.

The tangential velocity distribution in Fig.~\ref{fig:tangential_vel_total_fitting} and the best-fitting parameters on the right four panels of Fig.~\ref{fig:vel_dependence} are also informative.
Firstly, most of the LGAs with the smallest tangential velocity are coupled systems ($E_{\rm coupling} < 0$).
Secondly, coupled LGAs have typical tangential velocity $\lesssim 100\kms$, while uncoupled LGAs have a much wider distribution, extending to $\gtrsim 200\kms$ in low-density regions and $\gtrsim 300\kms$ in high-density regions.
Finally, the dispersion of the distribution is larger for LGAs with small separation.

In Figs.~\ref{fig:figure/radial_vel_fitting} and \ref{fig:tangential_vel_total_fitting}, we mark the locus of our LG using its measured radial and tangential velocities, shown as the vertical dashed lines in the middle-left panels, corresponding to its halo separation ($D\approx 519\hkpc$) and large-scale bias ($\tilde b\approx -2.7$).
Both of the radial and tangential velocities of our LG are consistent with those of analogue systems identified in the simulation, despite velocity information not being used in the LGA selection.

\section{Summary}%
\label{sec:summary}

Our LG is among the few galaxy systems for which detailed spatial kinematic, and structure properties can be characterised in great details, thus providing a powerful laboratory for testing $\Lambda$CDM cosmology.
At the same time, several properties of our LG have been highlighted as potential challenges to the $\Lambda$CDM, owing to the low probability of finding analogue systems to have similar characteristics in theoretical models.
However, such rarity estimates may be biased if they neglect the influence of the large-scale environment, which is known to play a crucial role in shaping the properties of galaxy and halo systems on small scales.
To address this, we investigate the connection between the halo pair configurations of LGA systems and the surround large-scale cosmic web using the ABACUSSUMMIT simulation suite.
Our results can be summarized as follows:
\begin{enumerate}

  \item
        We construct a LGA catalog using 25 ABACUSSUMMIT simulation boxes based on the criteria of two $10^{12\pm 0.3}\hMsun$ haloes separated by $400-700\hkpc$ and devoid of other massive neighbors.
        This results in $\approx 4.2\%$ of all haloes in the above mass range (see Figs.~\ref{fig:figure/demo} and \ref{fig:figure/lg_fraction_afo_exclusion_radius}).

  \item
        We show the joint distribution of coupling energy and specific orbital angular momentum for all LGAs.
        Our LG is enclosed within the 68 per cent contour line with typical $\|\mathbf j\|$ and $E_{\rm coupling}$ (see Fig.~\ref{fig:E_J_correlation}).

  \item
        The large-scale environment is quantified with the number overdensity of tracer haloes above $10^{11.6}\hMsun$ within $5-15\hmpc$.
        In addition, we prove that the mean number overdensity over a sample is in a linear relationship to the bias of that sample.
        Besides, we find that the redshift space distortion effect causes a constant shift in this linear relationship (see Fig.~\ref{fig:figure/halo_bias}).

  \item
        We show the joint distribution of coupling energy and large-scale bias for all LGAs, and our LG is located on the 95 per cent contour line with typical $E_{\rm coupling}$ and low $\tilde b$ (see Fig.~\ref{fig:figure/eb_env_joint_distribution}).

  \item
        We find significant correlation between LGA coupling energy and their large-scale overdensity.
        First, LGAs in over-dense regions have high fraction of uncoupled LGAs ($E_{\rm coupling} > 0$), where the fraction difference is up to $\approx 20\%$.
        Second, uncoupled LGAs live preferentially in more over-dense environments compared to their coupled counterparts ($E_{\rm coupling} < 0$).
        The ratio of the mean bias between these two samples of LGAs is $\approx 180\%$ (see Figs.~\ref{fig:lga_spatial_demo} and \ref{fig:figure/relation_eb_env}).

  \item
        The angular distribution of large-scale tracer haloes is found to be correlated to the spatial-kinematic configuration of LGAs.
        Moreover, the anisotropic pattern is controlled by the coupling energy: LGAs with low coupling energy have strong polar anisotropy and weak azimuthal anisotropy, while those with high coupling energy have weak polar anisotropy and strong azimuthal anisotropy (see Figs.~\ref{fig:healpix_logn} and \ref{fig:e_theta_phi}).

  \item
        We model the distribution of radial velocity and tangential velocity of LGA systems with student's $t$-distribution and gamma distribution, respectively.
        We also find the best-fitting parameters depend on the halo pair separation and large-scale environment (see Figs.~\ref{fig:figure/radial_vel_fitting}, \ref{fig:tangential_vel_total_fitting}, and \ref{fig:vel_dependence}).

\end{enumerate}

Although our LG serves as a powerful laboratory of testing $\Lambda$CDM model on non-linear scale, its statistical power is severely compromised due to the uniqueness of the system.
Our work finds that the LG lies within the 68 per cent contour line of all analogue systems identified in the $\Lambda$CDM simulation in the plane of coupling energy and specific orbital angular momentum.
By contrast, its large-scale overdensity is lower than that of 98.6 per cent of the analogue systems.
The LG is therefore typical in its orbital properties but atypical in its large-scale environment.
This work also highlights the crucial role of large-scale environment in shaping the spatial and kinematic properties of our LG, which also plays a non-negligible role in assessing the rarity of our LG system in comparison with model predictions.
Ongoing and future surveys, like DESI BGS \citep{hahnDESIBrightGalaxy2023} and 4MOST WAVES \citep{driver4MOSTConsortiumSurvey2019}, will extend this framework to observational samples, enabling direct comparison between simulated and observed Local Group analogues.

\section*{Acknowledgements}

KW thanks Isabel M. E. Santos-Santos, Alexander H. Riley, and Kyle Oman for helpful discussion.
KW \& PN acknowledges support from the Science and Technologies Facilities Council (STFC) through grant ST/X001075/1.
AF is supported by a Sweden’s Wallenberg Academy Fellowship.
LES is supported by the U.S. DOE Grant DE-SC0010813.

This work used the DiRAC@Durham facility managed by the Institute for Computational Cosmology on behalf of the STFC DiRAC HPC Facility (www.dirac.ac.uk).
The equipment was funded by BEIS capital funding via STFC capital grants ST/K00042X/1, ST/P002293/1, ST/R002371/1 and ST/S002502/1, Durham University and STFC operations grant ST/R000832/1.
DiRAC is part of the National e-Infrastructure.

The computation in this work is supported by the HPC toolkit \specialname[HIPP] \citep{hipp}, IPYTHON \citep{perezIPythonSystemInteractive2007}, MATPLOTLIB \citep{hunterMatplotlib2DGraphics2007}, NUMPY \citep{vanderwaltNumPyArrayStructure2011}, SCIPY \citep{virtanenSciPyFundamentalAlgorithms2020}, ASTROPY \citep{astropy:2022}.
Some of the results in this paper have been derived using the healpy and HEALPix package \citep{2005ApJ...622..759G, Zonca2019}.
This research made use of NASA’s Astrophysics Data System for bibliographic information.

\section*{Data availability}

The data underlying this article will be shared on reasonable request to the
corresponding author.

\bibliographystyle{mnras}
\bibliography{bibtex.bib}

\appendix

\section{Halo pair number distribution}
\label{sec:halo_pair_number_distribution}

\begin{figure}
  \begin{center}
    \includegraphics[width=\linewidth]{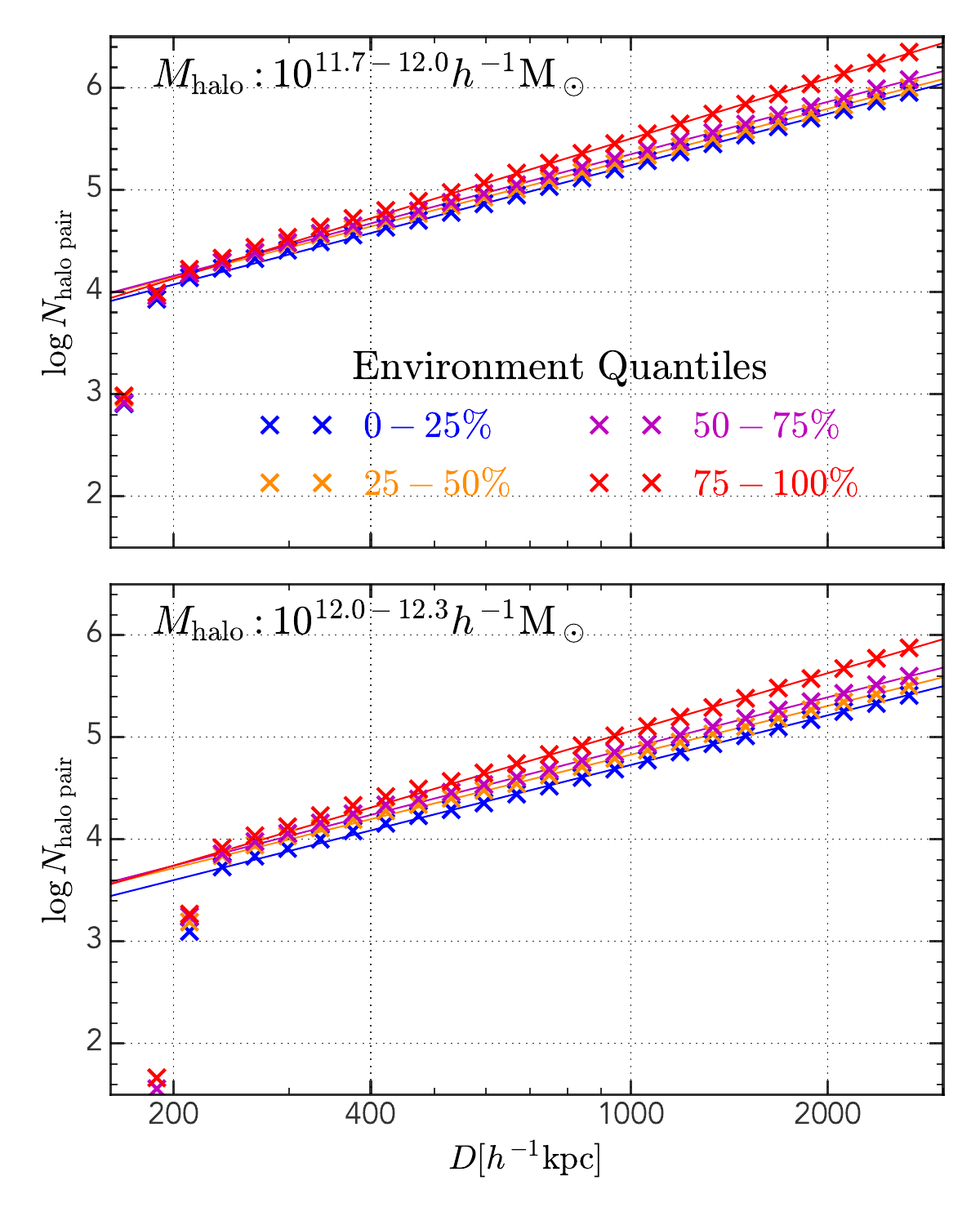}
  \end{center}
  \caption{
    The number of halo pairs as a function of separation in different halo mass bins and different environment bins, shown in coloured symbols.
    The solid lines with corresponding colours show the linear fitting in the logarithmic space using data points above $600h^{-1}\rm kpc$.
    This figure shows that the halo pair counts abruptly drops below some threshold distance due to the halo exclusion effect, but this threshold distance value is way below the separation ($400\hkpc$) in this work.
  }
  \label{fig:paircount_app}
\end{figure}

This work adopts the minimum halo pair separation of $400\hkpc$, which approaches the halo exclusion limit given that the virial radius of a $10^{12}\hMsun$ halo is about $200\hkpc$.
Haloes in simulations are often identified using the FoF algorithm, which can suffer from the ``bridging effect'', where two haloes are artificially linked by a few tenuously connected particles and treated as a single system.
Although \abacus uses the merger tree information to mitigate this issue \citep[see][]{boseConstructingHighfidelityHalo2022}, it remains necessary to verify that our LGA sample is not affected.

Fig.~\ref{fig:paircount_app} shows the halo pair number as a function of separation in different halo mass and environment bins.
The halo pair number increases linear in the logarithmic space above the threshold separation, below which the halo pair number drops sharply due to halo exclusion.
This threshold separation depends halo mass and environment, but remains around $200\hkpc$ in the mass range of interest, which is well below our adopted criterion of $400\hkpc$.
We therefore conclude that our LGA selection yields a complete halo pair sample.

\section{Uncertainty in the Local Group large-scale environment measurement}
\label{sec:uncertainty_mask}

\begin{figure}
  \begin{center}
    \includegraphics[width=0.95\linewidth]{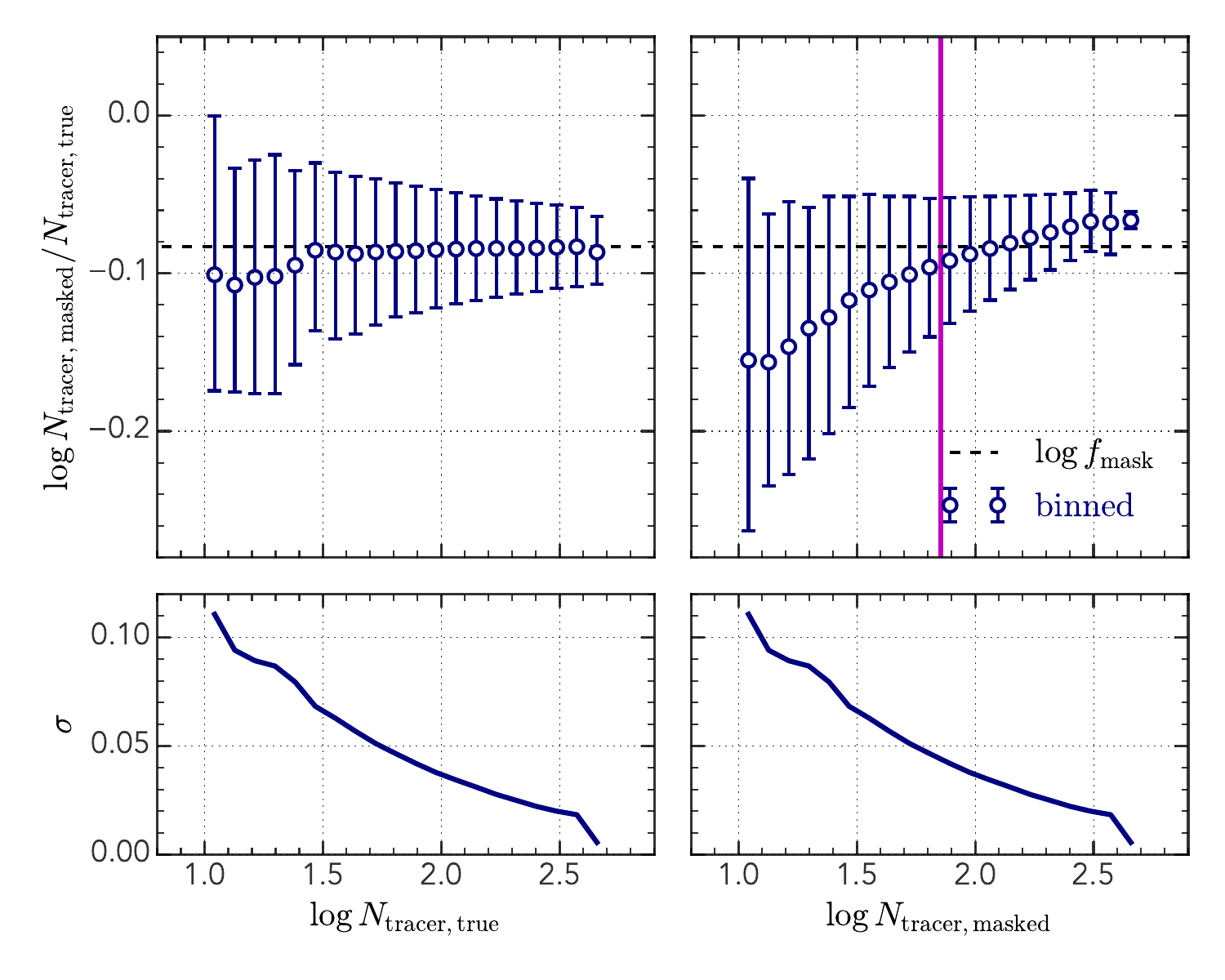}
  \end{center}
  \caption{
    \textbf{Top panels:} The mean tracer completeness ($\log N_{\rm tracer, masked}/N_{\rm tracer, true}$) as a function of true tracer numbers (left panel) and tracer numbers after masking (right panel), and the error bars show the 16\%-84\% percentiles.
    The navy horizontal dashed lines in both panel is the average completeness of all systems (1- 17.4\%=82.6\%).
    The vertical line shows the observed tracer halo number of our Local Group.
    \textbf{Bottom panels:} The navy lines are the standard deviation of tracer completeness.
    The red lines are the results of the Poisson process.
    The mean completeness is a constant value when expressed as a function of true tracer number, but biased when expressed as a function of masked tracer number.
  }
  \label{fig:N_tracer_uncertainty}
\end{figure}

The large-scale environment of our LG is estimated from the number overdensity of tracer haloes identified in the 2MRS survey.
Since the survey is incomplete due to extinction by the Milky Way, we mask a continuous band between $b=-10^\circ$ and $b=10^\circ$, corresponding to $\approx 17.4\%$ of the sky.
This masking introduces a bias in our large-scale environment estimation, and correcting for it requires knowledge of the full distribution of large-scale environments of LGAs.

To correct this systematic effect, we apply the same sky mask to all LGAs in the simulation, assuming the normal vector ($b=90^\circ$) is aligned with the $z$-axis of the simulation box.
We then count the number of tracer haloes after masking, denoted as $N_{\rm tracer, masked}$, and compare it with the true number of tracer haloes, $N_{\rm tracer, true}$, as shown in the top panels of Fig.~\ref{fig:N_tracer_uncertainty}.
Symbols with error bars show the mean and standard deviation of the completeness ($\log N_{\rm tracer, masked}/N_{\rm tracer, true}$), binned in $\log N_{\rm tracer, true}$ and $\log N_{\rm tracer, masked}$, respectively.
The horizontal black dashed line indicates the masked sky fraction, $\log f_{\rm mask} = \log (1 - 17.4\%)$.

When binned with $N_{\rm tracer, true}$, the mean completeness remains constant, confirming that the mask removes a fixed fraction of tracer haloes independent of the true tracer number.
However, when binned by the masked tracer number, which is the observable quantity, the relation becomes biased.
LGAs that originally contained many tracers but happen to lose more than average shift into low-$N_{\rm tracer, masked}$ bins, while those that lost fewer tracers move into high-$N_{\rm tracer, masked}$ bins.
Consequently, completeness falls below the mean at low $N_{\rm tracer, masked}$ end and rises above it at high-$N_{\rm tracer, masked}$ end.
Because observations only provide $N_{\rm tracer, masked}$, the top-right panel is used to infer the corresponding $N_{\rm tracer, true}$.

Finally, the number of tracer haloes around our LG in the 2MRS group catalogue, which is 59, is marked with the vertical solid line in the top-right panel.
This calibration implies a completeness of $\approx 10^{0.1}\approx 0.79$.
Applying this correction yields a total of $\approx 74.3\pm 8.1$ tracer haloes surrounding the MW, where the uncertainty is estimated from the bottom panel.

\bsp  
\label{lastpage}
\end{document}